\documentclass[11pt, onecolumn, journal]{IEEEtran}

\usepackage{multirow}
\usepackage{amsmath,amssymb,amscd}
\usepackage{graphicx,cite}

\newtheorem{lemma}{Lemma}
\newtheorem{rem}{Remark}

\begin{document}
\title{Beamforming Design for Joint Localization and Data Transmission in Distributed Antenna System{\Large $^{1}$}}

\author{\Large Seongah Jeong, Osvaldo Simeone, Alexander Haimovich and Joonhyuk Kang
\thanks{
Seongah Jeong and Joonhyuk Kang are with the Department of Electrical Engineering, Korea Advanced Institute of Science and Technology (KAIST), Daejeon, South Korea (Email: seongah@kaist.ac.kr and jhkang@ee.kaist.ac.kr).

Osvaldo Simeone and Alexander Haimovich are with the Center for Wireless Communications and Signal Processing Research (CWCSPR), ECE Department, New Jersey Institute of Technology (NJIT), Newark, NJ 07102, USA (Email: osvaldo.simeone@njit.edu and haimovic@njit.edu). 
}
}

\maketitle

\setcounter{footnote}{1}\footnotetext{This work was partially presented at IEEE GlobalSIP \cite{JSAGlobalSIP13}.}

\begin{abstract}
A distributed antenna system is studied whose goal is to provide data communication and positioning functionalities to Mobile Stations (MSs). Each MS receives data from a number of Base Stations (BSs), and uses the received signal not only to extract the information but also to determine its location. This is done based on Time of Arrival (TOA) or Time Difference of Arrival (TDOA) measurements, depending on the assumed synchronization conditions. The problem of minimizing the overall power expenditure of the BSs under data throughput and localization accuracy requirements is formulated with respect to the beamforming vectors used at the BSs. The analysis covers both frequency-flat and frequency-selective channels, and accounts also for robustness constraints in the presence of parameter uncertainty. The proposed algorithmic solutions are based on rank-relaxation and Difference-of-Convex (DC) programming.      
\end{abstract}

\section{Introduction}
Location-awareness is becoming an increasingly important feature of various wireless communication networks for security, disaster response and emergency relief, especially in GPS-denied environments \cite{Makela_02Mag, Kumar03spmag, Gezici_05spmag}. Examples include both commercial \cite{Makela_02Mag} and tactical networks \cite{Kumar03spmag}. In location-aware networks, fixed nodes, referred to as Base Stations (BSs), transmit/receive data over a radio interface to/from Mobile Stations (MSs), while at the same time serving also as anchors for the localization of the MSs. To this end, the positions of the BSs are known to all nodes, while the MSs attempt to estimate their positions based on the signals received from the BSs. Specifically, if BSs and MSs share a common time reference, localization can be performed via Time of Arrival (TOA) measurements \cite{Guvenc09CST, Shen12Globecom, Shen13ACM, Li13Arxiv, Wang13SigJ}, while, otherwise, Time Difference of Arrival (TDOA) \cite{Antique97VIRJIN}, Angle of Arrival \cite{Gezici_05spmag} or Received Signal Strength \cite{Gezici_05spmag} methods are known to be effective solutions. 

\begin{figure}[t]
\begin{center}
\includegraphics[width=10cm]{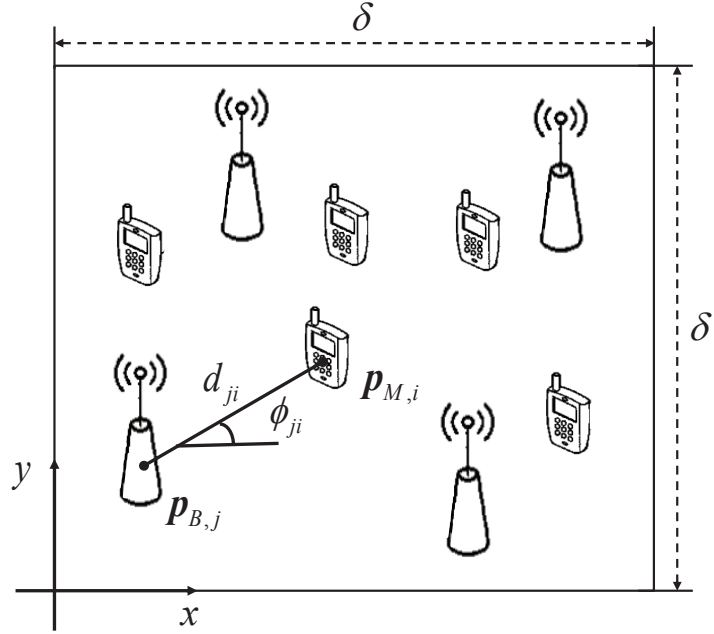}
\caption{The network under consideration consists of $N_B$ BSs with multiple antennas and $N_M$ single-antenna MSs. Each MS may receive data from all the BSs and uses the received pilot signals also to determine its position.} \label{fig:sys}
\end{center}
\end{figure}
The design of signal processing operations in a wireless network is conventionally targeted to account exclusively for communication-based performance criteria. For location-aware wireless networks, it is then relevant to revisit the conventional system design in order to accommodate also the localization requirements. Work along these lines can be found in \cite{Montal13ICASSP, Montal12ASIL, Larsen11ICASSP}. In \cite{Montal13ICASSP, Montal12ASIL, Larsen11ICASSP}, optimal pilot and data power allocation are investigated for a single BS in an orthogonal frequency division multiplexing (OFDM) system under constraints on the data rate and on the accuracy of TOA estimate. The latter criterion can be indirectly related to the localization precision \cite{Guvenc09CST, Shen12Globecom, Shen13ACM, Li13Arxiv, Wang13SigJ}. 

This paper considers the location-aware system in Fig. \ref{fig:sys}, in which multiple BSs, with multiple antennas, communicate with a number of single-antenna MSs. Each MS may receive data from all the BSs and uses the received signals also to estimate its location. The problem of interest is optimizing the beamforming vectors at the BSs, so as to minimize the overall power expenditure under data rate and localization accuracy constraints for all the MSs. The main contributions of this paper are as follows. 
\begin{itemize}
\item We first investigate beamforming optimization with rate and localization constraints under the assumption of frequency-flat channels and perfect knowledge of the system parameters, such as Channel State Information (CSI), at the BSs. Localization accuracy is measured by the Cram\'{e}r Rao Bound (CRB), and both TOA and TDOA-based positioning methods are considered, hence accounting for both synchronous and asynchronous set-ups (see, e.g., \cite{Kay93Book}). The proposed algorithms solve the resulting non-convex problems via rank-$1$ relaxation \cite{Luo10SPM, Beng99Allerton} and Difference-of-Convex (DC) programming \cite{Beck10Book}. 

\item A robust beamforming design strategy is proposed to combat the uncertainty on the system parameters at the BSs. The approach is based on a min-max formulation of the optimization problem (see, e.g., \cite{Shen12Globecom, Shen13ACM, Li13Arxiv, Ben09Book}). 

\item We extend the system design to frequency-selective channels under the assumption of OFDM transmission. In particular, a novel solution based on subcarrier grouping is proposed that is able to trade rate for localization accuracy. 

\item We provide extensive numerical results to assess the impact of the localization and data rate constraints including a case study concerning LTE-based system.
\end{itemize}

The rest of the paper is organized as follows. Sec. \ref{sec:sysmodel} presents the system model and formulates the problem of interest. In Sec. \ref{sec:beamforming}, we evaluate the localization metrics for the TOA and TDOA-based positioning methods and then describe the proposed beamforming strategies. In Sec. \ref{sec:robust}, a robust transmission strategy with respect to the uncertainty on the system parameters is proposed. Sec. \ref{sec:select} considers frequency-selective channels assuming OFDM transmission, and investigates corresponding optimal beamforming design problem. Finally, numerical results are given in Sec. \ref{sec:simul} and conclusions are drawn in Sec. \ref{sec:con}. 
    
\emph{Notation}:
$[\cdot]^T$ and $[\cdot]^*$ denote transpose and complex transpose, respectively; $|\pmb{A}|$ and tr$\{\pmb{A}\}$ are the determinant and the trace of a square matrix $\pmb{A}$, respectively; $[\cdot]_{n \times n}$ is the upper left $n \times n$ sub-matrix of its argument; $[\cdot]_{n,m}$ denotes element at the $n$th row and the $m$th column of its argument; $[\cdot]_{(a:b,c:d)}$ is the sub-matrix of its argument which corresponds to from the $a$th to the $b$th rows and from the $c$th to the $d$th columns; $\pmb{A} \succeq \pmb{B}$ means that matrix $\pmb{A}-\pmb{B}$ is positive-semidefinite; $||\pmb{x}||$ is the Euclidean norm of vector $\pmb{x}$; $\pmb{I}_n \in \mathrm{R}^{n \times n}$ is the identity matrix; $\pmb{0}_n$ and $\pmb{0}_{n \times m}$ are $n$-dimensional vector and $n \times m$ matrix of all zeros, respectively; $E[\cdot]$ denotes the expectation operator; $\lambda_{\rm{max}}(\pmb{A})$ and $\pmb{v}_{\rm{max}}(\pmb{A})$ are the maximum eigenvalue and corresponding eigenvector of a square matrix $\pmb{A}$, respectively.
\section{System Model}\label{sec:sysmodel}
The network under consideration is shown in Fig. \ref{fig:sys}, and it consists of $N_B$ BSs and $N_M$ MSs. The BS $j$ is equipped with $M_j$ antennas, while the MSs have a single antenna. The sets of BSs and MSs are denoted as $\mathcal{N}_B = \{1, 2, \dots, N_B\}$ and $\mathcal{N}_M=\{1, 2, \dots, N_M\}$, respectively. MS $i \in \mathcal{N}_M$ is located at position $\pmb{p}_{M,i}=[x_{M,i}\,y_{M,i}]^T \in \mathrm{R}^2$ in a $\delta \times \delta$ square area of the two-dimensional plane, while BS $j \in \mathcal{N}_B$ is located at position $\pmb{p}_{B,j}=[x_{B,j}\,y_{B,j}]^T \in \mathrm{R}^2$ in the same area. The distance and angle between BS $j$ and MS $i$ are defined as $d_{ji}=||\pmb{p}_{M,i}-\pmb{p}_{B,j}||$ and $\phi_{ji}=\tan^{-1}\frac{y_{M,i}-y_{B,j}}{x_{M,i}-x_{B,j}}$,
respectively (see Fig. \ref{fig:sys}). The positions $\pmb{p}_{B,j}$ of the BSs are known to all the nodes in the network. Each MS $i$ receives data from the BSs and uses the signals received from all the BSs also to estimate its location $\pmb{p}_{M,i}$. In order to make the localization possible, we assume $N_B \ge 3$ so that the position $\pmb{p}_{M,i}$ can be determined by each MS $i$ via triangulation based on time measurements from the received signals.
\subsection{Signal Model}\label{sec:sigmodel}
We start by detailing the system model for frequency-flat channels. Frequency-selective channels are treated in Sec. \ref{sec:select}. Each transmission block, of duration $T$, is divided into a training phase of $n_p$ symbols with total duration of $T_p$ and a phase for data transmission of length $n_d$ symbols with total duration of $T_d$. Throughout, unless stated otherwise, we will use the subscripts or superscripts $p$ and $d$ for variables related to pilots and data. Different BSs occupy orthogonal time-frequency resources, e.g., by using time-division multiple-access (TDMA) or frequency-division multiple-access (FDMA). Overall, in its dedicated resource, each BS $j$ transmits the signals 
\begin{subequations}\label{eq:x}
\begin{eqnarray}
\hspace{-0.9cm} \pmb{x}_j^{(p)}(t)&=&\sum_{i \in \mathcal{N}_M}\pmb{w}_{ji}s_{ji}^{(p)}(t)\label{eq:xp}\\
\hspace{-0.9cm} \text{and} \hspace{0.5cm}\pmb{x}_j^{(d)}(t)&=&\sum_{i \in \mathcal{N_M}}\pmb{w}_{ji}s_{ji}^{(d)}(t),\label{eq:xd}
\end{eqnarray}
\end{subequations}
in the training and data blocks, respectively, where $\pmb{w}_{ji}$ is the $M_j \times 1$ beamforming vector used for communication from the BS $j$ to the MS $i$, and 
\begin{subequations}\label{eq:s}
\begin{eqnarray}
s_{ji}^{(p)}(t)&=&\sum_{l=0}^{n_p-1}m_{ji}^{(p)}(l)g(t-lT_s)\label{eq:sp}\\
\text{and} \hspace{0.5cm} s_{ji}^{(d)}(t)&=&\sum_{l=0}^{n_d-1}m_{ji}^{(d)}(l)g(t-lT_s),\label{eq:sd} 
\end{eqnarray}
\end{subequations}
are the training and data signals used for communication between BS $j$ and MS $i$. In (\ref{eq:s}), $T_s$ is the symbol period; $g(t)$ is a (real) Nyquist pulse with unit energy, whose Fourier transform is $G(f)$; the pilot symbol sequences $m_{ji}^{(p)}(l)$, for $l=\left\{0,1,\dots,n_p-1\right\}$ with $i \in \mathcal{N}_M$, are orthogonal with unit amplitude and known to all nodes; and the data sequence $m_{ji}^{(d)}(l)$ for $l=\left\{0,1,\dots,n_d-1\right\}$ consists of the encoded data symbols from BS $j$ to MS $i$, which are assumed to be zero-mean independent random variable with correlation $E[m_{ji}^{(d)*}(l)m_{ji'}^{(d)}(l')]=\delta_{i-i'}\delta_{l-l'}$.

The channel between BS $j$ and MS $i$ is assumed here to be frequency-flat and constant within each transmission interval. Accordingly, the received signal at MS $i$ from BS $j$ during the entire training phase of duration $T_p$ can be written as 
\begin{equation}\label{eq:rxsig}
y_{ji}^{(p)}(t) = \zeta_{ji}\pmb{h}_{ji}^*\pmb{x}_j^{(p)}(t-\tau_{ji})+z_{ji}^{(p)}(t),
\end{equation}
where $\pmb{h}_{ji}$ is the $M_j \times 1$ complex channel vector between BS $j$ and MS $i$, which accounts for small-scale fading; $\tau_{ji}$ is the effective propagation delay between BS $j$ and MS $i$ given as
\begin{equation}\label{tau}
\tau_{ji}=\frac{d_{ji}}{c}+b_i,
\end{equation}
with $c$ being the propagation speed and $b_i$ being the time reference mismatch between the BSs and MS $i$ (see further discussion below); the noise $z_{ji}^{(p)}(t)$ is complex white Gaussian with zero mean and two-sided power spectral density $N_0$; and $\zeta_{ji}$ models the path loss between BS $j$ and MS $i$, which is given as 
\begin{equation}\label{pl}
\zeta_{ji}=\left(\frac{1}{1+\left(\frac{d_{ji}}{\Delta}\right)^\eta}\right)^{1/2},
\end{equation}
where $\eta$ is the path loss exponent and $\Delta$ is a reference distance  (see, e.g., \cite{Molit05Book}). The signal $y_{ji}^{(d)}(t)$ received during the data phase, of duration $T_d$, is similarly defined as
\begin{equation}
y_{ji}^{(d)}(t) = \zeta_{ji}\pmb{h}_{ji}^*\pmb{x}_j^{(d)}(t-\tau_{ji})+z_{ji}^{(d)}(t).
\end{equation}
Finally, we define the effective complex channel gain between BS $j$ and MS $i$ for the signal intended for MS $k$ as
\begin{equation}\label{eq:effgain}
\alpha_{ji}^{(k)}(\pmb{w}_{jk}) = \zeta_{ji}\pmb{h}_{ji}^*\pmb{w}_{jk}.
\end{equation} 
This definition is motivated by the fact that the received signal (\ref{eq:rxsig}), and similarly $y_{ji}^{(d)}(t)$, can be written as 
\begin{equation}
y_{ji}^{(p)}(t) = \sum_{k \in \mathcal{N}_M}\alpha_{ji}^{(k)}(\pmb{w}_{jk})s_{jk}^{(p)}(t-\tau_{ji})+z_{ji}^{(p)}(t),
\end{equation}
and hence the complex gain $\alpha_{ji}^{(k)}(\pmb{w}_{jk})$ affects the signal $s_{jk}^{(p)}(t)$ and $s_{jk}^{(d)}(t)$ transmitted by BS $j$ to MS $k$ at MS $i$.

Some further remarks are in order concerning the time references available at the MSs and the BSs in relation to the parameter $b_i$ in (\ref{tau}). The BSs are assumed to have a common time reference (e.g., via GPS). Instead, each MS $i$ has a time reference that is mismatched with respect to the common BSs' time reference by an offset $b_i$. This offset is generally unknown to the MSs and the BSs. We will first consider the case in which the offset $b_i$ is zero, which corresponds to a set-up where the MSs also have a common time reference with the BSs (e.g., via GPS) in Sec. \ref{sec:TOAspeb}. Then, in Sec. \ref{sec:TDOAspeb} we will cover the general case in which the  time reference mismatch $b_i$ is generally non-zero and unknown to all nodes. For this second case, we assume, for generality, that each MS $i$ has available some a priori knowledge about the offset $b_i$ in the form of a probability density function (pdf) $f(b_i)$.

In order to gain some initial insight into the problem, we first assume,  here and in Sec. \ref{sec:beamforming}, that the central unit that performs the optimization of the beamforming vectors knows the CSI $\zeta_{ji}\pmb{h}_{ji}$ for all $j \in \mathcal{N}_B$ and $i \in \mathcal{N}_M$ along with the inter-node distances $d_{ji}$ and the angles $\phi_{ji}$. The more practically relevant case with only imperfect CSI and parameter knowledge of the central unit is treated in Sec. \ref{sec:robust} building on the analysis in Sec. \ref{sec:beamforming}.
\subsection{Performance Metrics and Problem Formulation}\label{sec:perform}
The system design is concerned with guaranteeing acceptable performance both in terms of data transmission and of localization accuracy. These two requirements are discussed next.  
\subsubsection{Transmission Rate}\label{sec:achivrates}
Prior to decoding the data sequence $m_{ji}^{(d)}$ from the BS $j$, the MS $i$ performs timing recovery and channel estimation based on the received signal $y_{ji}^{(p)}(t)$ in (\ref{eq:rxsig}) during the training period $T_p$. While the maximum achievable rate generally depends on the specific channel estimate realization (see, e.g., \cite{Sam03TSP}), here we are interested in evaluating a measure of the achievable rate that can be calculated based on the CSI available at the BSs. This is in order to allow for the beamforming optimization at the central unit connected to the BSs. To this end, we write the transmission rate $r_{ji}(\pmb{W}_{j})$ (in bits/s/Hz) between BS $j$ and MS $i$ as
\begin{subequations}\label{eq:ratesflat}
\begin{eqnarray}
&&\hspace{-0.5cm}r_{ji}(\pmb{W}_{j}) = \frac{T_d}{N_BT}\log_2\left(1+\text{SINR}_{ji}(\pmb{W}_{j})\right),\label{eq:ratesflat}\\
&&\hspace{-1.6cm}\text{where}\nonumber\\
&&\hspace{-0.5cm}\text{SINR}_{ji}(\pmb{W}_{j})=\frac{\left|\alpha_{ji}^{(i)}(\pmb{w}_{ji})\right|^2}{N_0+\sum_{k \neq i, k \in \mathcal{N}_M}\left|\alpha_{ji}^{(k)}(\pmb{w}_{jk})\right|^2}
\end{eqnarray}
\end{subequations}
is the Signal-to-Interference-plus-Noise Ratio (SINR) between BS $j$ and MS $i$, and $\pmb{W}_j=\left\{\pmb{w}_{ji}\right\}_{i \in \mathcal{N}_M}$ collects all beamforming vectors of BS $j$. Note that, when calculating (\ref{eq:ratesflat}), we have assumed that each MS $i$ treats the interference coming from the undesired signals intended for the other MSs as additive noise{\footnotesize $^{2}$}\setcounter{footnote}{2}\footnotetext{A larger achievable rate could be achieved via a successive interference cancellation, but this is not further explored here.}. Moreover, the normalization by $N_B$ is due to the assumption of orthogonal transmissions by the BSs, which implies that each BS occupies only $1/N_B$ of the time-frequency resources. 
\subsubsection{Localization Accuracy}\label{sec:achivlocal}
Each MS $i$ estimates its location $\pmb{p}_{M,i}$ through the observation of the received signals (\ref{eq:rxsig}) from all the BS $j \in \mathcal{N}_B$ during the training period. In order to evaluate the localization accuracy, we adopt the squared position error (SPE) as the localization performance metric. This is defined for MS $i$ as (see e.g., \cite{Kay93Book,Shen10TIT})
\begin{equation}\label{eq:rho}
\rho_i(\pmb{W})=\text{E}\left[||\hat{\pmb{p}}_{M,i}-\pmb{p}_{M,i}||^2\right],
\end{equation} 
where $\hat{\pmb{p}}_{M,i}$ is the position estimate at MS $i$. We observe that the SPE $\rho_i(\pmb{W})$ depends on all beamforming vectors $\pmb{w}_{ji}$ for all $j \in \mathcal{N}_B$ and $i \in \mathcal{N}_M$, which are collectively denoted as $\pmb{W}=\left\{\pmb{W}_j\right\}_{j \in \mathcal{N}_B}$. 
\subsubsection{Problem Formulation}\label{sec:problem}
We denote $R_i$ and $Q_i$ as the rate and SPE localization requirements for the MS $i$, respectively. The problem of optimizing the beamforming vectors $\pmb{W}$ is then formulated as follows:
\begin{subequations}\label{eq:opt}
\begin{eqnarray}
&&{\mathop {\text{min} }\limits_{{\pmb{W}}} }  \hspace{0.7cm} {\sum_{j \in \mathcal{N}_B, \, i \in \mathcal{N}_M}\left\|\pmb{w}_{ji}\right\|^2}  \label{min}\\
&&\hspace{0.1cm}{\rm{s.t.}} \hspace{0.5cm}   {\sum_{j \in \mathcal{N}_B} r_{ji}(\pmb{W}_{j}) \ge R_i,} \label{ratereq}\\
&& \hspace{1.1cm}{\rho_i(\pmb{W}) \le Q_i,}  \hspace{0.5cm} \forall i \in \mathcal{N}_M. \label{localreq}
\end{eqnarray}
\end{subequations}
Note that the rate constraint (\ref{ratereq}) for the MS $i$ imposes that the total rate received from all BSs is larger than the required rate $R_i$. This constraint is appropriate if the BSs are connected to a common content delivery network and hence can all provide the required information to the MSs. This is for instance the case in distributed antenna systems \cite{Heath13Mag}. The localization constraint (\ref{localreq}) for the MS $i$ imposes that the SPE is smaller than the the required localization accuracy $Q_i$.   
\section{Beamforming Design}\label{sec:beamforming}
In this section, we first derive bounds on the SPE for TOA and TDOA-based localization. Then, using these bounds, we address the design of the beamforming vectors $\pmb{W}$ as per problem (\ref{eq:opt}).  
\subsection{Bounds on the SPE}\label{sec:local}
For any unbiased estimator of the position of MS $i$, the SPE can be bounded by the CRB as (see, e.g., \cite{Kay93Book})
\begin{equation}\label{SPEB}
\rho_i(\pmb{W}) \ge \text{tr}\left\{\pmb{J}_i^{-1}(\pmb{W})\right\},
\end{equation}
where $\pmb{J}_i(\pmb{W})$ is Equivalent Fisher Information Matrix (EFIM) for the estimation of the position $\pmb{p}_{M,i}$ (see e.g., \cite{Shen10TIT, Shen12Globecom, Shen13ACM, Li13Arxiv, Wang13SigJ}). The EFIM $\pmb{J}_i(\pmb{W})$ depends on whether the MS has a common time reference with the BSs or not. The first case, which can be modeled by setting $b_i=0$ in (\ref{tau}), is first discussed in Sec. \ref{sec:TOAspeb}, while the more general case is addressed in Sec. \ref{sec:TDOAspeb}.
\subsubsection{TOA-based Localization}\label{sec:TOAspeb}
We first assume the availability of a common time reference for MS $i$ and all BSs by setting $b_i=0$ in (\ref{tau}). In this case, localization can be performed by the MS $i$ through the estimation of the time delays, which are related to the BS-MS distance through (\ref{tau}), via triangulation. Hence, using conventional nomenclature, we refer to the localization under the assumption of a common time reference at the MS and BSs as being based on the estimation of the TOAs (see, e.g., \cite{Guvenc09CST, Shen12Globecom, Shen13ACM, Li13Arxiv, Wang13SigJ}). Under this assumption, as shown in Appendix \ref{app:flatTOA} following \cite{Shen10TIT}, the EFIM can be calculated as
\begin{eqnarray}\label{eq:TOAEFIM}
\hspace{-1cm}\pmb{J}_{i,\text{TOA}}(\pmb{W})&=&\frac{8\pi^2n_p\beta^2}{c^2}\sum_{j=1}^{N_B}\text{SNR}_{ji}(\pmb{W}_j)\pmb{q}_{ji}\pmb{q}_{ji}^T\nonumber\\
&=&\frac{8\pi^2n_p\beta^2}{c^2}\sum_{j=1}^{N_B}\text{SNR}_{ji}(\pmb{W}_j)\pmb{J}_\phi(\phi_{ji}),
\end{eqnarray} 
where $\pmb{q}_{ji}=[\cos\phi_{ji}\,\,\sin\phi_{ji}]^T$, $\beta$ is the effective bandwidth $\beta = \left\{\int_{-\infty}^{\infty}|fG(f)|^2df\right\}^{1/2}$
and we have defined the Signal-to-Noise Ratio (SNR) parameter $\text{SNR}_{ji}(\pmb{W}_j) = \sum_{k=1}^{N_M}\left|\alpha_{ji}^{(k)}(\pmb{w}_{jk})\right|^2/N_0$ along with the matrix
\begin{equation}\label{TOAJfi}
\pmb{J}_\phi(\phi)=\left[\begin{array}{cc}\cos^2\phi & \cos\phi\sin\phi \\ \cos\phi\sin\phi & \sin^2\phi \end{array}\right].
\end{equation}
\subsubsection{TDOA-based Localization}\label{sec:TDOAspeb}
We now consider the case where the time reference mismatch $b_i$ between the MS $i$ and BSs is possibly non-zero and is unknown to the MS and to the BSs. Due to the presence of this mismatch, the MSs cannot estimate directly the delays and hence TOA-based localization is not applicable. Instead, the classical approach in this case is to perform localization based on the estimate of the differences between the delays of all pairs of BSs. Therefore, we refer to localization in the presence of a MS-BSs time reference mismatch as being based on TDOAs \cite{Antique97VIRJIN}. 

From the a priori pdf $f(b_i)$, we can calculate the prior FIM $J_{b_i}=\text{E}_{b_i}\left\{\left[\frac{\partial\ln f(b_i)}{\partial b_i}\right]\left[\frac{\partial\ln f(b_i)}{\partial b_i}\right]^*\right\}$. The EFIM can be then calculated as
\begin{eqnarray}\label{eq:TDOAEFIMwb}
\pmb{J}_{i,\text{TDOA}}(\pmb{W})&=&\frac{8\pi^2n_p\beta^2}{c^2\left(\sum_{p=1}^{N_B}\text{SNR}_{pi}(\pmb{W}_p)+K_{b_i}\right)}\nonumber\\
&&\left\{K_{b_i}\sum_{m=1}^{N_B}\text{SNR}_{mi}(\pmb{W}_m)\pmb{J}_\phi(\phi_{mi})+\sum_{1 \le j < l \le N_B}\text{SNR}_{ji}(\pmb{W}_j)\text{SNR}_{li}(\pmb{W}_{l})\pmb{J}_\phi(\phi_{ji}, \phi_{li})\right\},
\end{eqnarray}
for all $p, m, l \in \mathcal{N}_B$, where $K_{b_i} =J_{b_i}/(8\pi^2n_p\beta^2)$ and we have defined the matrix 
\begin{equation}
\pmb{J}_\phi(\phi\hspace{-0.02cm}, \hspace{-0.02cm}\phi')\hspace{-0.1cm}=\hspace{-0.15cm}\left[\begin{array}{cc}\hspace{-2.1cm}(\cos\phi\hspace{-0.07cm}-\hspace{-0.07cm}\cos\phi')^{2} & \hspace{-1.7cm}(\cos\phi\hspace{-0.07cm}-\hspace{-0.07cm}\cos\phi')(\sin\phi\hspace{-0.07cm}-\hspace{-0.07cm}\sin\phi') \\ \hspace{-0.2cm}(\cos\phi\hspace{-0.07cm}-\hspace{-0.07cm}\cos\phi')(\sin\phi\hspace{-0.07cm}-\hspace{-0.07cm}\sin\phi') & \hspace{0.3cm}(\sin\phi\hspace{-0.07cm}-\hspace{-0.07cm}\sin\phi')^2 \end{array}\hspace{-0.2cm}\right]\hspace{-0.1cm}.
\end{equation}
The EFIM (\ref{eq:TDOAEFIMwb}) can be derived by following similar steps as in Appendix \ref{app:flatTOA}, which is derived in Appendix \ref{app:flatTDOA}.
\begin{rem}\label{rem1} 
In order to allow for an easier comparison of the localization accuracies achievable with TOA and TDOA-based localization, we can rewrite (\ref{eq:TDOAEFIMwb}) as 
\begin{equation}\label{eq:TDOAEFIMwb2}
\pmb{J}_{i,\text{TDOA}}(\pmb{W})=\pmb{J}_{i, \text{TOA}}(\pmb{W})-\frac{8\pi^2n_p\beta^2}{c^2\left(\sum_{p=1}^{N_B}\text{SNR}_{pi}(\pmb{W}_p)+K_{b_i}\right)}\left\{\sum_{j=1}^{N_B}\sum_{m=1}^{N_B}\text{SNR}_{ji}(\pmb{W}_j)\text{SNR}_{mi}(\pmb{W}_m)\pmb{q}_{ji}\pmb{q}_{mi}^T\right\},
\end{equation}
for all $p, m \in \mathcal{N}_B$. As expected, it can be seen that, if the time reference mismatch $b_i$ is perfectly known, i.e., if $J_{b_i}=\infty$ (or $K_{b_i}=\infty$), we have $\pmb{J}_{i,\text{TDOA}}(\pmb{W})=\pmb{J}_{i,\text{TOA}}(\pmb{W})$. 
\end{rem}
\subsection{Beamforming Design for TOA-based Localization}\label{sec:TOAbeam}
In this section, we elaborate on the solution of problem (\ref{eq:opt}) for TOA-based localization. We recall that the rate function $r_{ji}(\pmb{W}_j)$ in the constraint (\ref{ratereq}) is given as (\ref{eq:ratesflat}), while the SPE function $\rho_i(\pmb{W})$ in the constraint (\ref{localreq}) is bounded by the CRB (\ref{SPEB}) with EFIM (\ref{eq:TOAEFIM}).

Defining the covariance matrix $\pmb{\Sigma}_{ji}=\pmb{w}_{ji}\pmb{w}_{ji}^*$, problem (\ref{eq:opt}) can then be written as 
\begin{subequations}\label{opttoa}
\begin{eqnarray}
 &&\hspace{-1.4cm}{\mathop {\text{min} }\limits_{{\pmb{\Sigma}}} } {\hspace{1cm}\sum_{j \in \mathcal{N}_B, i \in \mathcal{N}_M}\text{tr}\left\{\pmb{\Sigma}_{ji}\right\}  }\label{eq:min}\\
 &&\hspace{-1.35cm}{\rm{s.t.}} { \hspace{0.04cm}\sum_{j \in \mathcal{N}_B}\hspace{-0.2cm} \frac{T_d}{N_BT}\log_2\hspace{-0.1cm}\left(\hspace{-0.1cm}1\hspace{-0.1cm}+\hspace{-0.1cm}\frac{\xi_{ji}^{(i)}(\pmb{\Sigma}_{ji})}{N_0\hspace{-0.1cm}+\hspace{-0.1cm}\sum_{k \neq i, k \in \mathcal{N}_M}\hspace{-0.1cm}\xi_{ji}^{(k)}(\pmb{\Sigma}_{jk})}\hspace{-0.1cm}\right)\hspace{-0.1cm} \ge \hspace{-0.1cm}R_i,} \label{eq:toaratereq}\\ 
&&{\hspace{-0.82cm}\text{tr}\hspace{-0.08cm}\left\{\hspace{-0.1cm}\left(\hspace{-0.1cm}\frac{8\pi^2n_p\beta^2}{c^2N_0}\hspace{-0.1cm}\sum_{j \in \mathcal{N}_B}\hspace{-0.05cm}\sum_{k \in \mathcal{N}_M}\hspace{-0.2cm}\xi_{ji}^{(k)}(\pmb{\Sigma}_{jk})\pmb{J}_\phi(\phi_{ji})\hspace{-0.1cm}\right)^{\hspace{-0.1cm}-1}\hspace{-0.05cm}\right\}\hspace{-0.1cm} \le \hspace{-0.1cm} Q_i,}   \label{eq:toaspebreq}\\ 
&& \hspace{-0.82cm}\text{rank}(\pmb{\Sigma}_{ji})=1\label{eq:rank},\\
&& {\hspace{-0.82cm}\pmb{\Sigma}_{ji} \succeq 0,} \hspace{0.2cm} \forall j \in \mathcal{N}_B \hspace{0.2cm} \text{and} \hspace{0.2cm} \forall i \in \mathcal{N}_M,\label{eq:psd}
\end{eqnarray}
\end{subequations}
where $\pmb{\Sigma}=\{\pmb{\Sigma}_{ji}\}_{j \in \mathcal{N}_B, i \in \mathcal{N}_M}$ and $\xi_{ji}^{(k)}(\pmb{\Sigma}_{jk})=\zeta_{ji}^2\pmb{h}_{ji}^*\pmb{\Sigma}_{jk}\pmb{h}_{ji}$ for $j \in \mathcal{N}_B$ and $i,k \in \mathcal{N}_M$. While the objective function is linear, problem (\ref{opttoa}) is complicated by the presence of the non-convex constraints (\ref{eq:toaratereq}) and (\ref{eq:rank}). Using rank-$1$ relaxation (see, e.g., \cite{Luo10SPM, Beng99Allerton}), and following the approach in \cite{Shen12Globecom, Li13Arxiv} to convert the localization constraint (\ref{eq:toaspebreq}) to a linear matrix inequality (LMI){\footnotesize $^{3}$}\setcounter{footnote}{3}\footnotetext{Using the approach in \cite{Shen12Globecom, Shen13ACM}, it is also possible to formulate the localization constraint (\ref{eq:toaspebreq}) as a second-order cone constraint.}, we propose the algorithm detailed in Algorithm $1$ for the solution of problem (\ref{opttoa}). The algorithm is based on the Majorization Minimization (MM) method for DC programming (see, e.g., \cite{Beck10Book}). Specifically, the algorithm first obtains a stationary point $\Sigma_{ji}^{\rm{opt}}$ for all $j \in \mathcal{N}_B$ and $i \in \mathcal{N}_M$ for the rank-relaxed problem (\ref{opttoa}) without the constraint (\ref{eq:rank}) using the MM algorithm, and then extracts a feasible solution for the original problem (\ref{opttoa}) using the standard rank-reduction approach (see, e.g., \cite{Luo10SPM, Beng99Allerton}). The details on the derivation of the algorithm and its properties can be found in Appendix \ref{app:TOAbeam}.
\begin{figure}
\begin{flushleft}
\begin{tabular}{|p{18cm}|}
\hline
$\bf{Algorithm}$ $\bf{1}$: Beamforming design for joint data transmission and TOA-based localization for frequency-flat channels\\
\hline
1. Initialize the matrices $\pmb{\Sigma}^{(1)}$ and $\pmb{M}_{i}$ to an arbitrary positive semidefinite matrices.\\ 
2. (\textit{MM algorithm}) Update the matrices $\pmb{\Sigma}^{(n+1)}$ as a solution of the following convex problem:
\begin{subequations}
\begin{eqnarray}\label{TOAlin}
&&\hspace{-0.85cm}{\mathop {\text{min} }\limits_{{\pmb{\Sigma}^{(n+1)},\,\, \pmb{M}_i}}}  \hspace{0cm}{\sum_{j \in \mathcal{N}_B, \, i \in \mathcal{N}_M}\hspace{-0.5cm}\text{tr}\left\{\pmb{\Sigma}_{ji}^{(n+1)}\right\}}  \\
&& \hspace{-0.8cm}{\rm{s.t.}}  \hspace{0.1cm} {R_i - \sum_{j \in \mathcal{N}_B}\hspace{-0.15cm}\frac{T_d}{N_BT} \hspace{-0.07cm}\log_2\hspace{-0.1cm}\left(\hspace{-0.1cm}N_0\hspace{-0.1cm}+\hspace{-0.2cm}\sum_{k \in \mathcal{N_M}}
             \hspace{-0.2cm}\xi_{ji}^{(k)}(\pmb{\Sigma}_{jk}^{(n+1)})\hspace{-0.1cm}\right)+\sum_{m \in \mathcal{N}_B}\frac{T_d}{N_BT}f(\pmb{\Sigma}_{m\sim i}^{(n+1)},\pmb{\Sigma}_{m\sim i}^{(n)} )\le 0,}\\
&& \hspace{-0.4cm} {  \left(\hspace{-0.25cm} \begin{array}{cc}
\pmb{M}_i & \hspace{-0.3cm}\pmb{I} \label{SchurTOA}\\
\pmb{I} &\hspace{-0.3cm}\frac{8\pi^2n_p\beta^2}{c^2N_0}\hspace{-0.1cm}\sum_{j \in \mathcal{N}_B}\hspace{-0.1cm}\sum_{k \in \mathcal{N}_M}\hspace{-0.1cm}\xi_{ji}^{(k)}\hspace{-0.05cm}(\pmb{\Sigma}_{jk}^{(n+1)})\pmb{J}_\phi(\phi_{ji}) \end{array}\hspace{-0.25cm} \right)\hspace{-0.1cm} \succeq \hspace{-0.05cm}0,}\\
&& \hspace{-0.4cm} \text{tr}\left\{\pmb{M}_i\right\} \le Q_i,\\
&& \hspace{-0.4cm}{\pmb{M}_i,\,\pmb{\Sigma}_{ji}^{(n+1)} \succeq 0,} \hspace{0.2cm} \forall j \in \mathcal{N}_B \hspace{0.2cm} \text{and} \hspace{0.2cm} \forall i \in \mathcal{N}_M,
\end{eqnarray}
\end{subequations}
where $\pmb{\Sigma}_{m\sim i}=\{\pmb{\Sigma}_{mk}\}_{k \neq i , k \in\mathcal{N}_M}$ and $f(\pmb{\Sigma}_{m\sim i}^{(n+1)},\pmb{\Sigma}_{m\sim i}^{(n)} )$ is a linear function defined as
\begin{equation}\label{mm}
f(\pmb{\Sigma}_{m\sim i}^{(n+1)},\pmb{\Sigma}_{m\sim i}^{(n)} )=\log_2\left(N_0+\sum_{l \neq i, \, l \in \mathcal{N_M}}\xi_{mi}^{(l)}(\pmb{\Sigma}_{ml}^{(n)})\right)+\frac{\sum_{p \neq i, \, p \in \mathcal{N_M}}\xi_{mi}^{(p)}(\pmb{\Sigma}_{mp}^{(n+1)})-\xi_{mi}^{(p)}(\pmb{\Sigma}_{mp}^{(n)})}{\ln2\left(N_0+\sum_{q \neq i, \, q \in \mathcal{N_M}}\xi_{mi}^{(q)}(\pmb{\Sigma}_{mq}^{(n)})\right)}. 
\end{equation}
3. Stop if $\sum_{j \in \mathcal{N}_B, \, i \in \mathcal{N}_M}\left\|\pmb{\Sigma}_{ji}^{(n+1)}-\pmb{\Sigma}_{ji}^{(n)}\right\|_F < \delta_{\rm{th}}$ with a predefined threshold value $\delta_{\rm{th}}$. Otherwise, $n \leftarrow n+1$ and go back to step 2.\\
4. (\textit{Rank reduction}) Extract the beamforming solution $\hat{\pmb{w}}_{ji}=\sqrt{\lambda_{\rm{max}}(\pmb{\Sigma}_{ji}^{\rm{opt}})}\pmb{v}_{\rm{max}}(\pmb{\Sigma}_{ji}^{\rm{opt}})$ from the optimal covariance matrix $\pmb{\Sigma}_{ji}^{\rm{opt}}$ obtained as the previous step for all $j \in \mathcal{N}_B$ and $i \in \mathcal{N}_M$.\\
5. Check whether the $\hat{\pmb{w}}_{ji}$ is feasible or not. If so, $\pmb{w}_{ji}^{\rm{opt}}=\hat{\pmb{w}}_{ji}$. Otherwise, rescale the $\hat{\pmb{w}}_{ji} \leftarrow (1+\delta_{\rm{inc}})\hat{\pmb{w}}_{ji}$ for any positive integer $\delta_{\rm{inc}}$ until $\hat{\pmb{w}}_{ji}$ is feasible.\\
\hline
\end{tabular}
\end{flushleft} 
\end{figure}
\subsection{Beamforming Design for TDOA-based Localization}\label{sec:TDOAbeam}
For TDOA-based localization, the only difference with respect to the TOA case treated above is the localization constraint (\ref{localreq}), where the EFIM matrix $\pmb{J}_{i,\text{TDOA}}$ in (\ref{eq:TDOAEFIMwb}) appears instead of $\pmb{J}_{i,\text{TOA}}$ in (\ref{eq:TOAEFIM}). Unlike $\pmb{J}_{i,\text{TOA}}$, the EFIM $\pmb{J}_{i,\text{TDOA}}$ is not linear over the covariance matrix $\pmb{\Sigma}_{ji}$. As a result, we cannot use the approach of \cite{Shen12Globecom, Li13Arxiv} employed in Appendix \ref{app:TOAbeam} to convert the localization accuracy constraint into a convex LMI.  

To deal with the problem identified above, we propose two different approaches. In the first approach, we observe that the following inequality between the EFIMs of TOA and TDOA-based localization holds:
\begin{equation}\label{LowerTDOA}
\pmb{J}_{i,\text{TDOA}}(\pmb{W}) \succeq \pmb{J}_{i,\text{TOA}}(\pmb{W})-\frac{8\pi^2n_p\beta^2N_M^2}{c^2K_{b_i}N_0^2}\sum_{j=1}^{N_B}\sum_{m=1}^{N_B}\zeta_{ji}^2\zeta_{mi}^2||\pmb{h}_{ji}||^2||\pmb{h}_{mi}||^2\pmb{q}_{ji}\pmb{q}_{mi}^T.
\end{equation}
This follows immediately from (\ref{eq:TDOAEFIMwb2}) using the inequalities $0 \le \text{SNR}_{ji}(\pmb{W}_j) \le N_M\zeta_{ji}^2||\pmb{h}_{ji}||^2/N_0$. Note that the bound (\ref{LowerTDOA}) is meaningful and tight  only when $K_{b_i}$ is sufficiently larger than $\sum_{j=1}^{N_B}\text{SNR}_{ji}(\pmb{W}_j)$. Since we have the inequality $\text{SNR}_{ji}(\pmb{W}_j) \le N_M\zeta_{ji}^2||\pmb{h}_{ji}||^2/N_0$ for any choice of $\pmb{W}_j$, this condition can be checked by comparing $K_{b_i}$ with $N_BN_M\zeta_{ji}^2||\pmb{h}_{ji}||^2/N_0$. Therefore, based on  (\ref{LowerTDOA}), we can obtain a feasible solution for the problem under study by solving problem (\ref{opttoa}) with
\begin{equation}\label{LowerTDOAconst} 
\frac{c^2N_0}{8\pi^2n_p\beta^2}\text{tr}\left\{\left(\sum_{j=1}^{N_B}\sum_{k=1}^{N_M}\xi_{ji}^{(k)}(\pmb{\Sigma}_{jk})\pmb{J}_\phi(\phi_{ji})-\frac{N_M^2}{K_{b_i}N_0}\sum_{l=1}^{N_B}\sum_{m=1}^{N_B}\zeta_{li}^2\zeta_{mi}^2||\pmb{h}_{li}||^2||\pmb{h}_{mi}||^2\pmb{q}_{li}\pmb{q}_{mi}^T\hspace{-0.1cm}\right)^{\hspace{-0.2cm}-1}\right\} \le Q_i
\end{equation}
in lieu of (\ref{eq:toaspebreq}). This problem can be addressed via Algorithm $1$, where we substitute (\ref{SchurTOA}) with the following constraint:
\begin{equation}\label{eq:TDOALMI}
\hspace{-0.1cm}\left( \begin{array}{ccc}
\hspace{-0.2cm}\pmb{M}_i &\hspace{-0.9cm}\pmb{I} & \\
\multirow{2}{*}{\hspace{-0.3cm}\it{{\pmb{I}}}}   & \hspace{-0.8cm}\frac{8\pi^2n_p\beta^2}{c^2N_0}\left(\sum_{j=1}^{N_B}\sum_{k=1}^{N_M}\xi_{ji}^{(k)}(\pmb{\Sigma}_{jk}^{(n+1)})\pmb{J}_\phi(\phi_{ji})\right.&\\
&\hspace{-0.4cm}-\left.\frac{N_M^2}{K_{b_i}N_0}\sum_{l=1}^{N_B}\sum_{m=1}^{N_B}\zeta_{li}^2\zeta_{mi}^2||\pmb{h}_{li}||^2||\pmb{h}_{mi}||^2\pmb{q}_{li}\pmb{q}_{mi}^T\right)& \end{array} \hspace{-0.6cm}\right) \succeq 0.
\end{equation}
As per Appendix \ref{app:TOAbeam}, this scheme provides a feasible solution, but is expected to be effective only when the prior information $J_{b_i}$ is sufficiently large, so that bound (\ref{LowerTDOA}) is tight. When this is not the case, we propose to use an alternative algorithm as described below. 

The idea behind the second proposed approach is to use a block coordinate iterative method, whereby the beamforming covariance matrices $\pmb{\Sigma}_j=\{\pmb{\Sigma}_{ji}\}_{i \in \mathcal{N}_M}$ of each BS $j$ are optimized in an iterative fashion over the BS index $j$ while fixing the other matrices $\pmb{\Sigma}_{j'i}$ for $j' \neq j$ and $j' \in \mathcal{N}_B$. The resulting algorithm, based on the MM approach, is detailed in Algorithm 2. The method uses the approximation of evaluating the denominator of (\ref{eq:TDOAEFIMwb}) for the localization constraint (\ref{localreq}) by its value obtained at the previous step (see (\ref{eq:TDOAEFIMwbR})). Similar to Algorithm $1$, Algorithm $2$ can be proved to always provide a feasible solution via scaling (see the steps in Algorithm $2$), as discussed in Appendix \ref{app:TOAbeam}. 
\begin{flushleft}
\begin{tabular}{|p{18cm}|}
\hline
$\bf{Algorithm}$ $\bf{2}$: Beamforming design for joint data transmission and TDOA-based localization for frequency-flat channels\\
\hline
1. Initialize the matrices $\pmb{\Sigma}^{(1)}$ and $\pmb{M}_{i}$ to an arbitrary positive semidefinite matrices.\\ 
2. (\textit{MM algorithm}) For $j=1:N_B$, successively update the matrices $\pmb{\Sigma}_j^{(n+1)}$ BS by BS as a solution of the following convex problem:
\begin{subequations}\label{opttdoa}
\begin{eqnarray}
&&\hspace{-0.85cm}{\mathop {\text{min} }\limits_{{\pmb{\Sigma}_j^{(n+1)},\,\, \pmb{M}_i}}} 
\hspace{0.1cm}{\sum_{i \in \mathcal{N}_M}\text{tr}\left\{\pmb{\Sigma}_{ji}^{(n+1)}\right\}}  \\
&&\hspace{-0.5cm}{\rm{s.t.}} \hspace{0.1cm} {R_i\hspace{-0.1cm}-\hspace{-0.3cm}\sum_{m \in \mathcal{N}_B}\hspace{-0.2cm}\frac{T_d}{N_BT} \hspace{-0.05cm}\log_2\hspace{-0.05cm}\left(N_0+\sum_{k \in \mathcal{N_M}}
             \xi_{mi}^{(k)}(h(\pmb{\Sigma}_{(j,m)k}^{(n+1)}))\right)+\sum_{p \in \mathcal{N}_B}\frac{T_d}{N_BT}f(h(\pmb{\Sigma}_{(j,p)\sim i}^{(n+1)}),\pmb{\Sigma}_{p\sim i}^{(n)} )\le 0,}\\
&& \hspace{0.05cm} {  \left( \begin{array}{cc}
\pmb{M}_i & \pmb{I} \\
\pmb{I} &  \pmb{J}_{i, \text{TDOA}}(\pmb{\Sigma}_{(j'\le j)}^{(n+1)},\pmb{\Sigma}^{(n)})\end{array} \right) \succeq 0,}\\
&& \hspace{0.05cm} \text{tr}\left\{\pmb{M}_i\right\} \le Q_i,\\
&& \hspace{0.05cm}{\pmb{M}_i,\,\pmb{\Sigma}_{ji}^{(n+1)} \succeq 0,} \hspace{0.5cm} \forall i \in \mathcal{N}_M,
 \end{eqnarray}
\end{subequations}
where $\pmb{\Sigma}_{(j' \le j)}^{(n+1)}=\left\{\pmb{\Sigma}_{j'i}^{(n+1)}\right\}_{j' \le j,\, j' \in \mathcal{N}_B \,\text{and}\,i \in \mathcal{N}_M}$, $h(\pmb{\Sigma}_{(j,j')i}^{(n+1)})=\left\{
\begin{array}{ll}
\pmb{\Sigma}_{j'i}^{(n+1)}, & j' \le j\\ 
\pmb{\Sigma}_{j'i}^{(n)}, & j' > j\\
\end{array}\right.$, $f(h(\pmb{\Sigma}_{(j,p)\sim i}^{(n+1)}),\pmb{\Sigma}_{p\sim i}^{(n)})$ is defined in (\ref{mm}) of Algorithm $1$ with $h(\pmb{\Sigma}_{(j,j')\sim i})=\{h(\pmb{\Sigma}_{(j,j')k})\}_{k \neq i, k \in \mathcal{N}_M}$ and 
\begin{eqnarray}\label{eq:TDOAEFIMwbR}
&&\hspace{-1cm}\pmb{J}_{i, \text{TDOA}}(\pmb{\Sigma}_{(j'\le j)}^{(n+1)},\pmb{\Sigma}^{(n)})=\frac{8\pi^2n_p\beta^2}{c^2N_0\left(\sum_{p \in \mathcal{N}_B}\sum_{q \in \mathcal{N}_M}\xi_{pi}^{(q)}(\pmb{\Sigma}_{pq}^{(n)})+K_{b_i}N_0\right)}\nonumber\\
&&\hspace{-0.8cm}\left\{K_{b_i}N_0\sum_{m \in \mathcal{N}_B}\sum_{k \in \mathcal{N}_M}\xi_{mi}^{(k)}(h(\pmb{\Sigma}_{(j,m)k}^{(n+1)}))\pmb{J}_\phi(\phi_{mi})+\hspace{-0.5cm}\sum_{1 \hspace{-0.02cm}\le \hspace{-0.02cm}l \hspace{-0.02cm}< \hspace{-0.02cm}t \hspace{-0.02cm}\le N_B}\hspace{-0.06cm}\sum_{u,v\in\mathcal{N}_M}\hspace{-0.06cm}\hspace{-0.2cm}\xi_{li}^{(u)}\hspace{-0.05cm}(\hspace{-0.03cm}h(\pmb{\Sigma}_{(j\hspace{-0.03cm},l)u}^{(n+1)}) \hspace{-0.03cm})\hspace{-0.02cm}\xi_{ti}^{(v)}(h(\pmb{\Sigma}_{(j\hspace{-0.03cm},t)v}^{(n+1)}) )\pmb{J}_\phi(\phi_{li}, \phi_{ti})\hspace{-0.1cm}\right\}\hspace{-0.1cm}.\nonumber
\end{eqnarray}
Note that for each BS $j$'s matrix $\pmb{\Sigma}_j^{(n+1)}$, stop if $\sum_{i \in \mathcal{N}_M}\left\|\pmb{\Sigma}_{ji}^{(n+1)}-\pmb{\Sigma}_{ji}^{(n)}\right\|_F < \delta_{\rm{th}}$ with a predefined threshold value $\delta_{\rm{th}}$.\\
3. Stop if the beamforming matrices $\pmb{\Sigma}^{(n+1)}$ are feasible solutions for all constraints. Otherwise, $n \leftarrow n+1$ and go back to step 2.\\
4. Extract $\pmb{w}_{ji}^{\rm{opt}}$ by following the steps $4$ and $5$ in Algorithm $1$.\\
\hline
\end{tabular}
\end{flushleft}   
\newpage
\section{Robust Beamforming Design}\label{sec:robust}
In practice, the inter-node distances $d_{ji}$, the angles $\phi_{ji}$ and the instantaneous CSI $\pmb{h}_{ji}$, for all $j \in \mathcal{N}_B$ and $i \in \mathcal{N}_M$ are not perfectly known to the central unit that performs the optimization of the beamforming vectors. Therefore, it is important to revisit the beamforming design discussed in the previous section by assuming that the mentioned parameters are only approximately available at the optimizer. Specially, as in \cite{Shen12Globecom, Shen13ACM, Li13Arxiv}, we assume that the inter-node distances $d_{ji}$ and the angles $\phi_{ji}$ are known within bounded uncertainty sets $S_{ji}^d$ and $S_{ji}^\phi$, respectively, as
\begin{subequations}\label{intervalpara}
\begin{eqnarray}
\hspace{-0.7cm}&&d_{ji} \in S_{ji}^d \triangleq [\hat{d}_{ji}-\epsilon_{ji}^d, \hat{d}_{ji}+\epsilon_{ji}^d],\label{intervald}\\
\hspace{-0.7cm} \text{and} &&\phi_{ji} \in S_{ji}^\phi \triangleq [\hat{\phi}_{ji}-\epsilon_{ji}^\phi, \hat{\phi}_{ji}+\epsilon_{ji}^\phi].\label{intervalang}
\end{eqnarray}
\end{subequations}         
In (\ref{intervalpara}), $\hat{d}_{ji}$ and $\hat{\phi}_{ji}$ are the nominal distance and angle parameters, and $\epsilon_{ji}^d$ and $\epsilon_{ji}^\phi$ are small positive numbers that define the uncertainty range for distances and angles, respectively. We observe that (\ref{intervald}) implies the uncertainty set for the path loss $\zeta_{ji} \in S_{ji}^\zeta \triangleq [\zeta_{ji}^{\rm{L}}, \zeta_{ji}^{\rm{U}}]$, where $\zeta_{ji}^{\rm{L}}=(1+(\frac{\hat{d}_{ji}+\epsilon_{ji}^d}{\Delta})^\eta)^{-1/2}$ and $\zeta_{ji}^{\rm{U}}=(1+(\frac{\hat{d}_{ji}-\epsilon_{ji}^d}{\Delta})^\eta)^{-1/2}$. As mentioned, the uncertainty model (\ref{intervalpara}) accounts for the lack of exact knowledge about an MS's position at the central unit. However, as it will be clarified by the problem formulation given below, it may also be used to represent the presence of an arbitrary number of MSs with distances and angles as in (\ref{intervalpara}), all of which have the same localization requirement and wish to receive the same information. As for the CSI, we assume that the central unit is aware only of the second-order statistics $\pmb{R}_{ji}=E[\pmb{h}_{ji}\pmb{h}_{ji}^*]$. Since the second-order statistics of the channel depend on the MS position (see Sec. \ref{sec:simul}), this choice is again appropriate for both the scenarios with a single MS with uncertain position and with multiple MSs within the uncertainty region (\ref{intervalpara}) with common localization and rate requirements.

We propose to formulate the optimization problem by adopting a min-max robust approach (see \cite{Ben09Book}) as follows:
\begin{subequations}\label{eq:Ropt}
\begin{eqnarray}
&&\hspace{-1.7cm}{\mathop {\text{min} }\limits_{{\pmb{W}}} \hspace{-0.1cm}\mathop {\text{max} }\limits_{\left\{\zeta_{ji} \in S_{ji}^\zeta, \,\phi_{ji} \in S_{ji}^\phi\right\}}} 
{\sum_{j \in \mathcal{N}_B, \, i \in \mathcal{N}_M}\left\|\pmb{w}_{ji}\right\|^2}  \\
&&\hspace{-0.9cm}{\rm{s.t.}} \hspace{0.9cm}   {\sum_{j \in \mathcal{N}_B} \bar{r}_{ji}(\pmb{W}_{j}, \zeta_{ji}) \ge R_i,} \label{Rratereq}\\
&& \hspace{0.5cm}{\text{tr}\left\{\bar{\pmb{J}_i}^{-1}(\pmb{W}, \zeta_{i}, \phi_{i})\right\} \le Q_i,}  \hspace{0.2cm} \forall i \in \mathcal{N}_M, \label{Rlocalreq}
\end{eqnarray}
\end{subequations}
where $\zeta_i=\{\zeta_{ji}\}_{j \in \mathcal{N}_B}$ and  $\phi_i=\{\phi_{ji}\}_{j \in \mathcal{N}_B}$.
In problem (\ref{eq:Ropt}), in order to make rates and EFIMs computable based only on the available second-order statistics $\pmb{R}_{ji}$, we defined the achievable data rate $\bar{r}_{ji}(\pmb{W}_{j}, \zeta_{ji})$ and EFIM $\bar{\pmb{J}_{i}}(\pmb{W}, \zeta_{i}, \phi_{i})$ for both TOA and TDOA-based localization by substituting $|\alpha_{ji}^{(k)}(\pmb{w}_{jk})|^2$ with its expectation $\bar{\xi}_{ji}^{(k)}(\pmb{w}_{jk},\zeta_{ji})=\zeta_{ji}^2\pmb{w}_{jk}^*\pmb{R}_{ji}\pmb{w}_{jk}$ for $j \in \mathcal{N}_B$ and $i,k \in \mathcal{N}_M$. This leads to 
\begin{subequations}
\begin{equation}
\bar{r}_{ji}(\pmb{W}_j, \zeta_{ji})\hspace{-0.1cm}=\hspace{-0.1cm}\frac{T_d}{N_BT}\log_2\hspace{-0.1cm}\left(\hspace{-0.1cm}1\hspace{-0.1cm}+\hspace{-0.05cm}\frac{\bar{\xi}_{ji}^{(i)}(\pmb{w}_{ji}, \zeta_{ji})}{N_0\hspace{-0.1cm}+\hspace{-0.1cm}\sum_{k \neq i, k \in \mathcal{N}_M}\hspace{-0.1cm}\bar{\xi}_{ji}^{(k)}(\pmb{w}_{jk}, \zeta_{ji})}\hspace{-0.1cm}\right)\hspace{-0.05cm},\label{Rrate}
\end{equation}
\begin{equation}
\bar{\pmb{J}}_{i,\text{TOA}}(\pmb{W}, \zeta_{i}, \phi_{i})\hspace{-0.1cm}=\hspace{-0.15cm}\sum_{j=1}^{N_B}\sum_{k=1}^{N_M}\hspace{-0.05cm}\frac{8\pi^2n_p\beta^2}{c^2N_0}\bar{\xi}_{ji}^{(k)}(\pmb{w}_{jk},\zeta_{ji})\pmb{J}_\phi(\phi_{ji})\label{RefimTOA}\hspace{0.5cm}
\end{equation}
and
\begin{eqnarray}
&&\hspace{-0.7cm}\bar{\pmb{J}}_{i,\text{TDOA}}(\pmb{W}\hspace{-0.05cm},\hspace{-0.05cm} \zeta_{i},\hspace{-0.05cm} \phi_{i} \hspace{-0.05cm})\hspace{-0.1cm}=\hspace{-0.1cm}
\frac{8\pi^2n_p\beta^2}{c^2N_0\hspace{-0.05cm}\left(\hspace{-0.05cm}\sum_{p=1}^{N_B}\sum_{q=1}^{N_M}\hspace{-0.05cm}\bar{\xi}_{pi}^{(q)}(\pmb{w}_{pq},\zeta_{pi})\hspace{-0.1cm}+\hspace{-0.1cm}K_{b_i}N_0\hspace{-0.05cm}\right)}\nonumber\\
&&\hspace{-0.7cm}\left\{K_{b_i}N_0\sum_{m=1}^{N_B}\sum_{k=1}^{N_M}\bar{\xi}_{mi}^{(k)}(\pmb{w}_{mk},\zeta_{mi})\pmb{J}_\phi(\phi_{mi})+\hspace{-0.2cm}\sum_{1 \le l < t \le N_B}\hspace{-0.05cm}\sum_{u=1}^{N_M}\sum_{v=1}^{N_M}\bar{\xi}_{li}^{(u)}(\pmb{w}_{lu},\zeta_{li})\bar{\xi}_{ti}^{(v)}(\pmb{w}_{tv},\zeta_{ti}) \pmb{J}_\phi(\phi_{li}, \phi_{ti})\right\}\hspace{-0.1cm},\label{RefimTDOAwPri}
\end{eqnarray}
\end{subequations}
for all $p,m,l,t \in \mathcal{N}_B$. In the following, we solve this problem for TOA and TDOA-based localization. 
\subsection{TOA-based Localization}\label{sec:TOArobust}
We first observe that any solution of problem (\ref{eq:Ropt}) with (\ref{Rrate}) and the TOA EFIM (\ref{RefimTOA}) for the constraint (\ref{Rratereq}) and (\ref{Rlocalreq}), respectively, must have $d_{ji}^*=\hat{d}_{ji}+\epsilon_{ji}^d$ and hence $\zeta_{ji}^*=\zeta_{ji}^{\rm{L}}$. In other words, the worst-case distance $d_{ji}^*$ is the largest distance in the uncertainty set $S_{ji}^d$. To see this, it is sufficient to note that the achievable data rate of each MS, namely $\bar{r}_{ji}(\pmb{W}_j, \zeta_{ji})$ in (\ref{Rratereq}), is a monotonically non-decreasing function of $\zeta_{ji}$, and hence non-increasing function of $d_{ji}$, and so is the CRB $\text{tr}\{\bar{\pmb{J}}_{i,\text{TOA}}^{-1}(\pmb{W}, \zeta_{i}, \phi_{i})\}$. 

In contrast, the maximization over the angles $\phi_{ji}$ is only relevant to the localization constraint (\ref{Rlocalreq}) and is not a convex problem, which makes it difficult to obtain a closed-form solution for $\phi_{ji}$.  Instead, in order to find the worst-case angle $\phi_{ji}^*$ in the uncertainty set $S_{ji}^\phi$, we adopt the relaxation method proposed in \cite{Shen12Globecom,Li13Arxiv}, whereby the matrix $\pmb{J}_\phi(\phi_{ji})$ in the EFIM (\ref{RefimTOA}) is substituted with 
\begin{equation}\label{QfiTOA}
\pmb{Q}_{\phi}(\hat{\phi}_{ji})=\pmb{J}_\phi(\hat{\phi}_{ji})-\sin\epsilon_{ji}^\phi\pmb{I}.
\end{equation} 
As shown in \cite{Shen12Globecom,Li13Arxiv}, the matrix $\pmb{Q}_{\phi}(\hat{\phi}_{ji})$ guarantees that the following inequality relationship 
\begin{equation}\label{Jyrel}
\text{tr}\hspace{-0.05cm}\left\{\hspace{-0.08cm}\bar{\pmb{J}}_{i,\text{TOA}}^{-1}(\pmb{W}, \zeta_{i}, \phi_{i})\hspace{-0.08cm}\right\}\hspace{-0.1cm} \le \hspace{-0.1cm}\text{tr}\hspace{-0.05cm}\left\{\hspace{-0.08cm}\bar{\pmb{Q}}_{i,\text{TOA}}^{-1}(\pmb{W}, \zeta_{i})\hspace{-0.08cm}\right\}\hspace{-0.1cm} 
\le \hspace{-0.1cm}\text{tr}\hspace{-0.05cm}\left\{\hspace{-0.08cm}\bar{\pmb{Q}}_{i,\text{TOA}}^{-1}(\pmb{W}, \zeta_{i}^*)\hspace{-0.08cm}\right\}
\end{equation}
holds for all $\zeta_{ji} \in S_{ji}^\zeta$ and $\phi_{ji} \in S_{ji}^\phi$, where $\zeta_i^*=\{\zeta_{ji}^*\}_{j \in \mathcal{N}_B}$ and $\bar{\pmb{Q}}_{i,\text{TOA}}(\pmb{W}, \zeta_{i})=\sum_{j=1}^{N_B}\sum_{k=1}^{N_M}\frac{8\pi^2n_p\beta^2}{c^2N_0}$ $\bar{\xi}_{ji}^{(k)}(\pmb{w}_{jk},\zeta_{ji})$ $\pmb{Q}_{\phi}(\hat{\phi}_{ji})$. These inequalities ensure that the rightmost side of (\ref{Jyrel}) provides a conservative measure of the SPE for all positions within the uncertainty set (\ref{intervalpara}). 

Given the discussion above, the robust optimization problem (\ref{eq:Ropt}) is reformulated as
\begin{subequations}\label{eq:RoptTOA}
\begin{eqnarray}
{\mathop {\text{min} }\limits_{{\pmb{W}}} } \hspace{0.5cm}&&  {\hspace{-0.6cm}\sum_{j \in \mathcal{N}_B, \, i \in \mathcal{N}_M}\left\|\pmb{w}_{ji}\right\|^2}  \label{minR}\\
{\rm{s.t.}} \hspace{0.5cm} &&  {\hspace{-0.6cm}\sum_{j \in \mathcal{N}_B} \bar{r}_{ji}(\pmb{W}_j, \zeta_{ji}^*) \ge R_i,} \label{RrateTOAreq}\\
&& {\hspace{-0.6cm}\text{tr}\left\{\bar{\pmb{Q}}_{i,\text{TOA}}^{-1}(\pmb{W}, \zeta_{i}^*)\right\} \le Q_i,}  \hspace{0.3cm} \forall i \in \mathcal{N}_M. \label{RlocalTOAreq}
\end{eqnarray}
\end{subequations}    
We propose to resolve the problem (\ref{eq:RoptTOA}) by using the rank-1 relaxation combined with the MM algorithm as done for problem (\ref{eq:opt}) in Sec. \ref{sec:TOAbeam}. The detailed algorithm can be easily derived as for Algorithm $1$ and is not reported here. 
\subsection{TDOA-based Localization}\label{sec:TDOArobust}
Consider now problem (\ref{eq:Ropt}) with the TDOA EFIM (\ref{RefimTDOAwPri}) for the localization constraint (\ref{Rlocalreq}). Since the CRB $\text{tr}\{\bar{\pmb{J}}_{i,\text{TDOA}}^{-1}(\pmb{W}, \zeta_{i}, \phi_{i})\}$ is a monotonically non-decreasing function of $\zeta_{ji}$, like in the TOA case, $d_{ji}^*=\hat{d}_{ji}+\epsilon_{ji}^d$ is the solution of problem (\ref{eq:Ropt}) and therefore $\zeta_{ji}^*=\zeta_{ji}^{\rm{L}}$. Moreover, the maximization over the angles $\phi_{ji}$ is not a convex problem. To cope with this issue, similar to the TOA case, we obtain a universal upper bound on the CRB that holds for all $\phi_{ji} \in S_{ji}^\phi$ and $\phi_{j'i} \in S_{j'i}^\phi$. This bound is akin to (\ref{QfiTOA}) derived in \cite{Li13Arxiv} and is based on the matrix 
\begin{equation}\label{QfiTDOA}
\pmb{Q}_{\phi}(\hat{\phi}_{ji}, \hat{\phi}_{j'i})=\pmb{J}_\phi(\hat{\phi}_{ji}, \hat{\phi}_{j'i})-\left(\sin\epsilon_{ji}^\phi+\sin\epsilon_{j'i}^\phi+4\sin\left(\frac{\epsilon_{ji}^\phi+\epsilon_{j'i}^\phi}{2}\right)\right)\pmb{I}.
\end{equation} 
The following Lemma $\ref{lem1}$ summarizes the main conclusion of the analysis.
\begin{lemma}\label{lem1}
If $\bar{\pmb{Q}}_{i,\text{TDOA}}(\pmb{W}, \zeta_{i}) \succeq 0$, the following inequality holds for all $\zeta_{ji} \in S_{ji}^\zeta$ and $\phi_{ji} \in S_{ji}^\phi$: 
\begin{equation}\label{QfiTDOArel}
\text{tr}\hspace{-0.05cm}\left\{\hspace{-0.08cm}\bar{\pmb{J}}_{i\hspace{-0.05cm},\text{TDOA}}^{-1}(\pmb{W}\hspace{-0.05cm},\hspace{-0.05cm} \zeta_{i}\hspace{-0.05cm}, \hspace{-0.05cm}\phi_{i})\hspace{-0.08cm}\right\}\hspace{-0.15cm} \le \hspace{-0.1cm}\text{tr}\hspace{-0.05cm}\left\{\hspace{-0.08cm}\bar{\pmb{Q}}_{i\hspace{-0.05cm},\text{TDOA}}^{-1}(\pmb{W}\hspace{-0.05cm},\hspace{-0.05cm} \zeta_{i})\hspace{-0.08cm}\right\}\hspace{-0.15cm} \le \hspace{-0.1cm}\text{tr}\hspace{-0.05cm}\left\{\hspace{-0.08cm}\bar{\pmb{Q}}_{i\hspace{-0.05cm},\text{TDOA}}^{-1}(\pmb{W}\hspace{-0.05cm}, \hspace{-0.05cm}\zeta_{i}^*)\hspace{-0.08cm}\right\}\hspace{-0.1cm},
\end{equation}
where
\begin{eqnarray}
&&\hspace{-0.7cm}\bar{\pmb{Q}}_{i,\text{TDOA}}(\hspace{-0.05cm}\pmb{W}\hspace{-0.05cm}, \hspace{-0.05cm}\zeta_{i}\hspace{-0.05cm})\hspace{-0.05cm}=
\hspace{-0.05cm}\frac{8\pi^2n_p\beta^2}{c^2N_0\left(\hspace{-0.05cm}\sum_{p=1}^{N_B}\sum_{q=1}^{N_M}\bar{\xi}_{pi}^{(q)}(\pmb{w}_{pq},\zeta_{pi}\hspace{-0.05cm})\hspace{-0.05cm}+\hspace{-0.05cm}K_{b_i}N_0\hspace{-0.05cm}\right)}\nonumber\\
&&\hspace{-0.5cm}\left\{K_{b_i}N_0\sum_{m=1}^{N_B}\sum_{k=1}^{N_M}\bar{\xi}_{mi}^{(k)}(\pmb{w}_{mk},\zeta_{mi})\pmb{Q}_{\phi}(\hat{\phi}_{mi})+\hspace{-0.35cm}\sum_{1 \le l < t \le N_B}\hspace{-0.05cm}\sum_{u=1}^{N_M}\hspace{-0.05cm}\sum_{v=1}^{N_M}\bar{\xi}_{li}^{(u)}(\pmb{w}_{lu},\zeta_{li})\bar{\xi}_{ti}^{(v)}(\pmb{w}_{tv},\zeta_{ti}) \pmb{Q}_{\phi}(\hat{\phi}_{li}, \hat{\phi}_{ti})\hspace{-0.1cm}\right\}\hspace{-0.1cm},
\end{eqnarray}
for all $p,m,l,t \in \mathcal{N}_B$.
\end{lemma}
\begin{IEEEproof}
The proof is in Appendix \ref{app:QfiTDOA}.
\end{IEEEproof}
Based on Lemma \ref{lem1}, the robust optimization problem (\ref{eq:Ropt}) for TDOA-based localization can be written as in (\ref{eq:RoptTOA}) with the constraint $\text{tr}\left\{\bar{\pmb{Q}}_{i,\text{TDOA}}^{-1}(\pmb{W}, \zeta_{i}^*)\right\} \le Q_i$ in lieu of (\ref{RlocalTOAreq}). We propose to address this problem by a block coordinate iterative method similar to the technique introduced in Sec. \ref{sec:TDOAbeam}. The details can be easily derived based on the discussion therein and are not reported here.  
\section{Frequency-Selective Fading Channels}\label{sec:select}
We now turn to the investigation of system operating over frequency-selective fading channels via OFDM transmission. We first detail the system model in Sec. \ref{sec:sigmodelselect}. We then formulate the problem of beamforming optimization in Sec. \ref{sec:performselect} and propose an algorithm for its solution in Sec. \ref{sec:beamselect}. Throughout, we focus on a TOA-based localization assuming perfect knowledge of the system parameters. Extensions to TDOA-based localization and robust optimization can be performed in a similar fashion discussed in the previous section and are left to future work. 
\subsection{Signal Model}\label{sec:sigmodelselect}
\begin{figure}[t]
\begin{center}
\includegraphics[width=10cm]{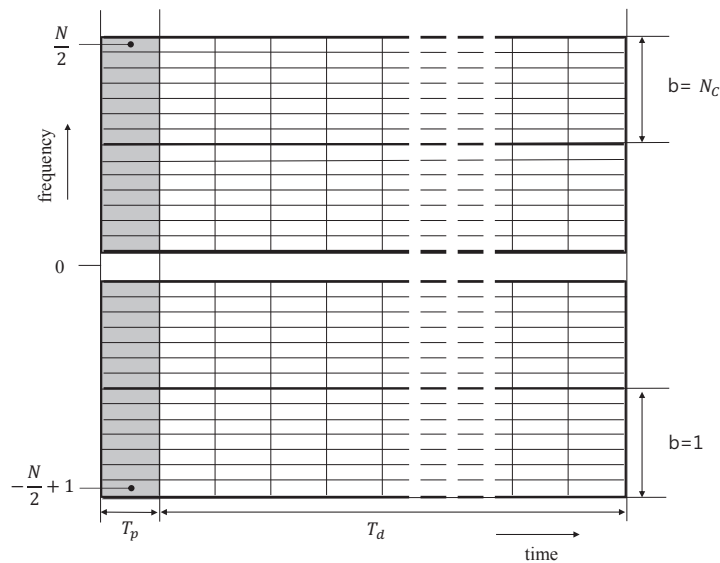}
\caption{The time-frequency structure of the transmitted signal. Shaded subcarriers contain pilot symbols.} \label{fig:ofdmsig}
\end{center}
\end{figure}

We assume OFDM transmission with $N$ subcarriers, which is taken to be even for simplicity. We assume that the duration of the cyclic prefix is larger than the channel delay spread plus the time delay uncertainty, which ensures the zero inter-block interference (see e.g., \cite{Molit05Book}). We denote the pilot or data symbols transmitted from BS $j$ to MS $i$ at a subcarrier $n$ as $S_{ji,n}^{(p)}$ or $S_{ji,n}^{(d)}$, respectively. The pilot symbols have unit energy, i.e., $|S_{ji,n}^{(p)}|=1$, and the encoded data symbols are zero-mean independent random variables with $E[S_{ji,n}^{(d)*}S_{ji',n'}^{(d)}]=\delta_{i-i'}\delta_{n-n'}$. In accordance with various wireless standards, one pilot OFDM signal is followed by $T_d$ data OFDM symbols, as shown in Fig. \ref{fig:ofdmsig}. The sampling period is $T_s$, and the bandwidth is $1/T_s$. 

We consider the following standard multi-path frequency-selective channel model between BS $j$ and MS $i$ (see e.g., \cite{Larsen11ICASSP, Montal12ASIL, Montal13ICASSP, Wang13SigJ})
\begin{equation}
\pmb{h}_{ji}(t)=\sum_{l=0}^{L_i-1}\pmb{h}_{ji,l}\delta(t-lT_s-\tau_{ji}),
\end{equation}
where $L_i$ is a known upper bound on the number of discrete multi-path components between all BSs and MS $i$; $\tau_{ji}$ is the delay between BS $j$ and MS $i$ and $\pmb{h}_{ji,l}$ is the $M_j \times 1$ complex channel vector accounting for the spatial response of the $l$th path. For all channels between BS $j$ and MS $i$, we define the $M_j \times L_i$ channel matrix given as $\pmb{H}_{ji}=[\pmb{h}_{ji,0}\,\,\cdots\,\, \pmb{h}_{ji,L_i-1}]$.

We allow the same beamforming vector $\pmb{w}_{ji,b}$ to be used in each $b$th block of $N/N_C$ subcarriers as illustrated in Fig. \ref{fig:ofdmsig}, where $N_C$ is the number of blocks. As it will be discussed below, the number of subcarriers in each block should be larger than the number of multi-paths, i.e., $N/N_C > L_i$, in order to enable localization (see Remark \ref{rem2}). The $N/N_C \times 1$ vector containing the received signal at the $b$th block during the pilot phase is given by (see e.g., \cite{Larsen11ICASSP, Montal12ASIL, Montal13ICASSP})
\begin{equation}\label{ofdmpilot}
\pmb{Y}_{ji,b}^{(p)}=\sum_{k=1}^{N_M}\pmb{S}_{jk,b}^{(p)}\pmb{\Gamma}_{b}(\tau_{ji})\pmb{F}_{L_i,b}\pmb{\alpha}_{ji,b}^{(k)}(\pmb{w}_{jk,b})+\pmb{Z}_{ji,b}^{(p)}
\end{equation}
for $b \in \mathcal{N}_C=\{1,2,\dots,N_C\}$, where we define $c_{bn}=\frac{N}{N_C}(b-1)-\frac{N}{2}+n$ as the $n$th subcarrier in the $b$th block; $\pmb{S}_{jk,b}^{(p)}=\text{diag}\{S_{jk,c_{b1}}^{(p)},$ $\dots, S_{jk,c_{bN/N_C}}^{(p)}\}$ collects the pilot signals transmitted by BS $j$ to MS $k$ in block $b$;  $\pmb{\Gamma}_b(\tau_{ji})=\text{diag}\{-\exp\{j\frac{2\pi}{T_s}$ $c_{b1}\tau_{ji}\},\dots,\exp\{-j\frac{2\pi}{T_s}c_{bN/N_C}\tau_{ji}\}\}$ accounts for the contribution of the delay $\tau_{ji}$; $\alpha_{ji,lb}^{(k)}(\pmb{w}_{jk,b})=\zeta_{ji}\pmb{h}_{ji,l}^*\pmb{w}_{jk,b}$ is the effective channel gain by the transmission for MS $k$ on the $l$th path between BS $j$ and MS $i$ for $j \in \mathcal{N}_B$ and $i,k \in \mathcal{N}_M$ (cf. (\ref{eq:effgain})); $\pmb{\alpha}_{ji,b}^{(k)}(\pmb{w}_{jk,b})=[\alpha_{ji,0b}^{(k)}(\pmb{w}_{jk,b})\,\,\cdots\,\,$ $\alpha_{ji,(L_i-1)b}^{(k)}(\pmb{w}_{jk,b})]^T$; $\pmb{Z}_{ji,b}^{(p)}$ is the additive white Gaussian noise with power $N_0$; and $\pmb{F}_{L_i,b}$ is the $b$th $N/N_C \times L_i$ matrix
\begin{equation}
\pmb{F}_{L_i,b}\hspace{-0.1cm}=\hspace{-0.15cm}\left[\begin{array}{cccc}
1 &e^{-j\frac{2\pi}{N}c_{b1}}&\cdots & e^{-j\frac{2\pi}{N}c_{b1}(L_i-1)}\\
\vdots&\vdots&&\vdots\\
1&e^{-j\frac{2\pi}{N}c_{bN/N_C}}&\cdots&e^{-j\frac{2\pi}{N}c_{bN/N_C}(L_i-1)}
\end{array}\right].
\end{equation}
The received signal $\pmb{Y}_{ji,b}^{(d)}$ during the data phase is similarly defined. 
\subsection{Performance Metrics and Problem Formulation}\label{sec:performselect}
As throughout the paper, we are interested in minimizing the power expenditure under data rate and localization accuracy constraints. 
\subsubsection{Transmission Rate}\label{sec:achivratesselect}
Treating the interference as additive noise, the achievable transmission rate $r_{ji}(\pmb{W}_j)$ (in bits/s/Hz) between BS $j$ and MS $i$ is given by (cf. (\ref{eq:ratesflat}))
\begin{subequations}
\begin{equation}
r_{ji}(\pmb{W}_{j})\hspace{-0.1cm} = \hspace{-0.1cm}\frac{T_d}{N_B\hspace{-0.05cm}(T_d\hspace{-0.05cm}+\hspace{-0.1cm}1\hspace{-0.05cm})}\hspace{-0.05cm}\sum_{b=1}^{N_C}\hspace{-0.1cm}\sum_{n=1}^{N/N_C}\hspace{-0.15cm}\log_2\left(1+\text{SINR}_{ji,bn}(\pmb{W}_{j,b})\right),\label{eq:ratesselect}
\end{equation}
where
\begin{eqnarray}
&&\hspace{-0.65cm}\text{SINR}_{ji,bn}(\pmb{W}_{j,b})=\hspace{-0.1cm}\frac{\left|\sum_{l=0}^{L_i-1}\hspace{-0.1cm}\alpha_{ji,lb}^{(i)}(\hspace{-0.03cm}\pmb{w}_{ji,b}\hspace{-0.03cm})e^{-j\frac{2\pi}{N}c_{bn}l}\right|^2}{N_0\hspace{-0.1cm}+\hspace{-0.1cm}\sum_{k \neq i, k \in \mathcal{N}_M}\hspace{-0.05cm}\left|\sum_{l=0}^{L_i-1}\hspace{-0.1cm}\alpha_{ji,lb}^{(k)}(\pmb{w}_{jk,b})e^{-j\frac{2\pi}{N}c_{bn}l}\right|^2} 
\end{eqnarray}
\end{subequations}
is the SINR between BS $j$ and MS $i$ at the $n$th subcarrier in the $b$th block, and we set $\pmb{W}_{j,b}=\left\{\pmb{w}_{ji,b}\right\}_{i \in \mathcal{N}_M}$ and $\pmb{W}_j=\left\{\pmb{W}_{j,b}\right\}_{b \in \mathcal{N}_c}$.  
\subsubsection{Localization Accuracy}\label{sec:achivlocalselect}
Focusing on the case with TOA-based localization, i.e., with $b_i=0$ in (\ref{tau}), 
the EFIM of the MS $i$'s position can be calculated as (cf. (\ref{eq:TOAEFIM}))
\begin{subequations}
\begin{eqnarray}
&&\hspace{-0.65cm}\pmb{J}_{i,\text{TOA}}(\pmb{W})=\frac{8\pi^2}{c^2T_s^2}\sum_{j=1}^{N_B}\sum_{b=1}^{N_C}\text{SNR}_{ji,b}(\pmb{W}_{j,b})\pmb{J}_\phi(\phi_{ji}),\label{FIMTOAselec}\\
&&\hspace{-4cm}\text{where}\nonumber\\
&&\hspace{-0.65cm}\text{SNR}_{ji,b}(\pmb{W}_{j,b})=\hspace{-0.1cm}\frac{1}{N_0}\hspace{-0.1cm}\sum_{k \in \mathcal{N}_M}\hspace{-0.2cm}\pmb{\alpha}_{ji,b}^{(k)}(\hspace{-0.05cm}\pmb{w}_{jk,b}\hspace{-0.05cm})^*\hspace{-0.05cm}\pmb{F}_{L_i,b}^*\pmb{K}_b\pmb{\Pi}^{\bot}_{\pmb{F}_{L_i,b}}\pmb{K}_b\pmb{F}_{L_i,b}\pmb{\alpha}_{ji,b}^{(k)}
(\hspace{-0.05cm}\pmb{w}_{jk,b}\hspace{-0.05cm})\hspace{-0.02cm},
\end{eqnarray}
\end{subequations}
with $\pmb{\Pi}^{\bot}_{\pmb{F}_{L_i,b}}=\pmb{I}-\pmb{F}_{L_i,b}$ $(\pmb{F}_{L_i,b}^*\pmb{F}_{L_i,b})^{-1}\pmb{F}_{L_i,b}^*$ being the orthogonal projection onto the orthogonal complement of the column space of $\pmb{F}_{L_i,b}$ and $\pmb{K}_b=\text{diag}\{c_{b1},\dots,c_{bN/N_C}\}$. The derivation is similar to Appendix \ref{app:flatspeb} and is derived in Appendix \ref{app:ofdmTOA}.
\begin{rem}\label{rem2}
The condition of $N/N_C > L_i$ for all $i \in \mathcal{N}_M$ is necessary in order to enable localization. In fact, if $N/N_C \le L_i$, the EFIM in (\ref{FIMTOAselec}) is singular. 
\end{rem}
\subsubsection{Problem Formulation}\label{sec:problemselect}
With (\ref{eq:ratesselect}) and (\ref{SPEB}) with (\ref{FIMTOAselec}) for the rate and localization constraints, respectively, the problem of optimizing all the beamforming vectors $\pmb{W}=\{\pmb{W}_j\}_{j \in \mathcal{N}_B}$ can be now formulated as (\ref{eq:opt}), where the objective function (\ref{min}) is substituted with $\sum_{j \in \mathcal{N}_B, \, i \in \mathcal{N}_M, b \in \mathcal{N}_C}\left\|\pmb{w}_{ji,b}\right\|^2$.

\subsection{Beamforming Design}\label{sec:beamselect}
We propose an approach to solve the optimization problem of Sec. \ref{sec:problemselect} that follows the method introduced in Sec. \ref{sec:TOAbeam} and derived in Appendix \ref{app:TOAbeam}. The details of the algorithm can be found in Algorithm $3$.  
\begin{figure}
\begin{flushleft}
\begin{tabular}{|p{18cm}|}
\hline
$\bf{Algorithm}$ $\bf{3}$: Beamforming design for joint data transmission and TOA-based localization for frequency-selective channels\\
\hline
1. Initialize the matrices $\pmb{\Sigma}^{(1)}=\{\pmb{\Sigma}_{ji,b}^{(1)}\}_{j \in \mathcal{N}_B\,i \in \mathcal{N}_M\,\text{and}\, b \in \mathcal{N}_C}$ and $\pmb{M}_{i}$ to an arbitrary positive semidefinite matrices, where $\pmb{\Sigma}_{ji,b}=\pmb{w}_{ji,b}\pmb{w}_{ji,b}^*$.\\ 
2. (\textit{MM algorithm}) Update the matrices $\pmb{\Sigma}^{(n+1)}$ as a solution of the following convex problem:
\begin{subequations}\label{opttoaselect}
\begin{eqnarray}
&&\hspace{-0.75cm}{\mathop {\text{min} }\limits_{{\pmb{\Sigma}^{(n+1)},\,\, \pmb{M}_i}}} \hspace{0cm}{\sum_{j \in \mathcal{N}_B, i \in \mathcal{N}_M, b \in \mathcal{N}_C}\hspace{-0.2cm}\text{tr}\left\{\pmb{\Sigma}_{ji,b}^{(n+1)}\right\}}\label{eq:minselect}\\
&&\hspace{-0.15cm}{\rm{s.t.}}\nonumber\\
&&\hspace{-0.75cm} R_i\hspace{-0.08cm}-\hspace{-0.25cm}\sum_{j \hspace{-0.03cm}\in\hspace{-0.02cm} \mathcal{N}_B}\hspace{-0.06cm}\sum_{b \hspace{-0.02cm}\in \mathcal{N}_C}\hspace{-0.12cm}\sum_{n_c\hspace{-0.04cm}=\hspace{-0.02cm}1}^{N/N_C}\hspace{-0.2cm}\frac{T_d}{N_B\hspace{-0.03cm}(T_d\hspace{-0.06cm}+\hspace{-0.1cm}1)}\hspace{-0.04cm}\log_2\hspace{-0.1cm}\left(\hspace{-0.13cm} N_0\hspace{-0.09cm}+\hspace{-0.3cm}\sum_{k \in \mathcal{N}_M} \hspace{-0.2cm}\xi_{ji\hspace{-0.03cm},bn_c}^{\hspace{-0.02cm}(k)}\hspace{-0.06cm}(\hspace{-0.03cm}\pmb{\Sigma}_{jk\hspace{-0.03cm},b}^{(n+1)}\hspace{-0.04cm})\hspace{-0.1cm}\right)+\hspace{-0.2cm}\sum_{p \in \mathcal{N}_B}\hspace{-0.05cm}\sum_{b \in \mathcal{N}_C}\hspace{-0.1cm}\sum_{n_c=1}^{N/N_C}\frac{T_d}{N_B(T_d+1)}g(\pmb{\Sigma}_{p\sim i,b}^{(n+1)},\pmb{\Sigma}_{p\sim i,b}^{(n)} )\le 0, \label{eq:toaratereqselect}\nonumber\\
&&\\ 
&&\hspace{-0.77cm} {\left( \hspace{-0.27cm}\begin{array}{cc}
\pmb{M}_i & \hspace{-0.4cm}\pmb{I} \\
\hspace{-0.15cm}\pmb{I} & \hspace{-0.35cm}\frac{8\pi^2}{c^2T_s^2N_0}\hspace{-0.1cm}\sum_{j \hspace{-0.02cm}\in \mathcal{N}_B}\hspace{-0.1cm}\sum_{k \hspace{-0.02cm}\in \mathcal{N}_M}\hspace{-0.1cm}\sum_{b \hspace{-0.02cm}\in \mathcal{N}_C}\hspace{-0.12cm}\xi_{ji\hspace{-0.02cm},b}^{\hspace{-0.02cm}(k)}\hspace{-0.03cm}(\pmb{\Sigma}_{jk\hspace{-0.02cm},b}^{\hspace{-0.02cm}(n+1)}\hspace{-0.03cm})\pmb{J}_\phi(\hspace{-0.02cm}\phi_{ji}\hspace{-0.02cm}) \end{array}\hspace{-0.3cm} \right)\hspace{-0.15cm} \succeq \hspace{-0.05cm}0,}\label{eq:toalocalreqselect}\\
&& \hspace{-0.5cm} \text{tr}\left\{\pmb{M}_i\right\} \le Q_i,  \\
&&\hspace{-0.5cm}{\pmb{M}_i,\,\pmb{\Sigma}_{ji,b}^{(n+1)} \succeq 0,} \hspace{0.2cm} \forall j \in \mathcal{N}_B \hspace{0.1cm} i \in \mathcal{N}_M \hspace{0.1cm}\text{and} \hspace{0.1cm} \forall b \in \mathcal{N}_C, \label{eq:psdselect}
 \end{eqnarray}
\end{subequations}
where $\pmb{f}_{ji,bn_c}=\sum_{l=0}^{L_i-1}\pmb{h}_{ji,l}\exp\{-j\frac{2\pi}{N}c_{bn_c}l\}$, $\xi_{ji,bn_c}^{(k)}(\pmb{\Sigma}_{jk,b})=\zeta_{ji}^2\pmb{f}_{ji,bn_c}^*\pmb{\Sigma}_{jk,b}\pmb{f}_{ji,bn_c}$, 
$\xi_{ji,b}^{(k)}(\pmb{\Sigma}_{jk,b})=\text{tr}\{\zeta_{ji}^2\pmb{H}_{ji}\pmb{F}_{L_i,b}^*\pmb{K}_b\pmb{\Pi}^{\bot}_{\pmb{F}_{L_i,b}}\pmb{K}_b\pmb{F}_{L_i,b}\pmb{H}_{ji}^*\pmb{\Sigma}_{jk,b}\}$  , $\pmb{\Sigma}_{j\sim i,b}=\{\pmb{\Sigma}_{jk,b}\}_{k \neq i, k \in \mathcal{N}_M}$ and $g(\pmb{\Sigma}_{p\sim i,b}^{(n+1)},\pmb{\Sigma}_{p\sim i,b}^{(n)} )$ is a linear function defined as (cf. (\ref{mm}))
\begin{eqnarray}\label{mm2}
&&\hspace{-1cm}g(\pmb{\Sigma}_{p\sim i,b}^{(n+1)},\pmb{\Sigma}_{p\sim i,b}^{(n)} )\hspace{-0.1cm}=\hspace{-0.05cm}\log_2\hspace{-0.05cm}\left(\hspace{-0.1cm}N_0+\sum_{q \neq i, \, q \in \mathcal{N_M}}
             \xi_{pi,bn_c}^{(q)}(\pmb{\Sigma}_{pq,b}^{(n)})\right)+\frac{\sum_{s \neq i, \, s \in \mathcal{N_M}}
             \left(\xi_{pi,bn_c}^{(s)}(\pmb{\Sigma}_{ps,b}^{(n+1)})-\xi_{pi,bn_c}^{(s)}(\pmb{\Sigma}_{ps,b}^{(n)})\right)}{\ln2\left(N_0+\sum_{t \neq i, \, t \in \mathcal{N_M}}
             \xi_{pi,bn_c}^{(t)}(\pmb{\Sigma}_{pt,b}^{(n)})\right)}. 
\end{eqnarray}
3. Stop if $\sum_{j \in \mathcal{N}_B, \, i \in \mathcal{N}_M,\, b \in \mathcal{N}_C}\left\|\pmb{\Sigma}_{ji,b}^{(n+1)}-\pmb{\Sigma}_{ji,b}^{(n)}\right\|_F < \delta_{\rm{th}}$ with a predefined threshold value $\delta_{\rm{th}}$. Otherwise, $n \leftarrow n+1$ and go back to step 2.\\
4. Extract $\pmb{w}_{ji,b}^{\rm{opt}}$ by following the steps $4$ and $5$ in Algorithm $1$.\\
\hline
\end{tabular}
\end{flushleft} 
\end{figure}
\newpage
\section{Numerical Results}\label{sec:simul}
\begin{figure}[t]
\begin{center}
\includegraphics[width=6cm]{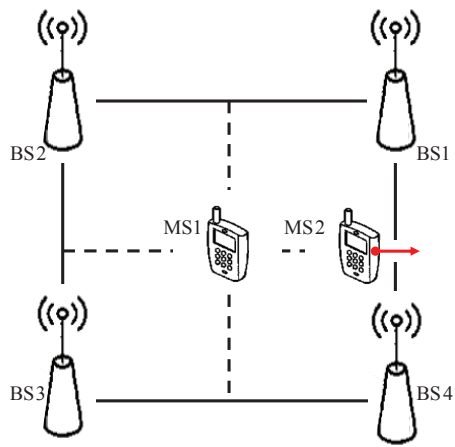}
\caption{Set-up for the numerical results in Fig. \ref{fig2}.} \label{figset}
\end{center}
\end{figure}
\begin{figure}[t]
\begin{center}
\includegraphics[width=8cm]{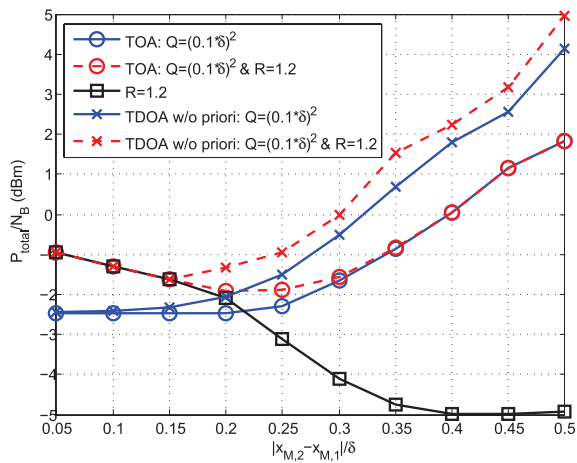}
\caption{Normalized per-BS transmit power as a function of MS 2's position for $(M, N_B, N_M) = (4, 4, 2)$. $N_B=4$ BSs are placed at the corners of the square region, while MS $1$ is located in the center, and MS $2$ moves on the x-axis away from MS $1$ (see Fig. \ref{figset}).} \label{fig2}
\end{center}
\end{figure}
In this section, we evaluate the performance of the proposed beamforming strategies for frequency-flat and then for frequency-selective channels. We also consider a case study using LTE-based system parameters. Unless stated otherwise, the size of the area is $\delta=200$m and the reference distance $\Delta$ in (\ref{pl}) is chosen so that the path loss at distance $100$m is $\zeta^2=-110$dB. Moreover, we assume noise level $N_0=-121$dBm and a path loss exponent of $\eta=4$. Identical requirements for data rate and localization accuracy of $R$ and $Q$ are applied to all MSs.
\subsection{Frequency-Flat Fading Channels}\label{sec:simulflat}
We assume training and data phases with $n_p=10$, $T_d/T=2/3$ and an effective bandwidth of $\beta=200$kHz. The frequency-flat channel between BS $j$ and MS $i$ is modeled as $\pmb{h}_{ji}=\pmb{s}_{ji}(\phi_{ji})$ with steering vector $\pmb{s}_{ji}(\phi_{ji})=[1 \,\,e^{j\pi\cos\phi_{ji}}\,\,\cdots\,\,e^{j\pi(M-1)\cos\phi_{ji}}]$ for $j \in \mathcal{N}_B$ and $i \in \mathcal{N}_M$, hence assuming that each BS is equipped with an $M$-element uniform linear antenna array with half-wavelength spacing. We optimize the beamforming vectors using Algorithm $1$ for TOA-based localization and using the best between Algorithm $2$ and the first method based on the (\ref{LowerTDOA}) discussed in Sec. \ref{sec:TDOAbeam} for TDOA-based localization.

In Fig. \ref{fig2}, we consider a network with $(M, N_B, N_M) = (4, 4, 2)$, where $N_B=4$ BSs are placed at the vertices of the square region of Fig. \ref{fig:sys}, while MS 1 is located in the center of square area, and MS $2$ moves on the x-axis away from MS $1$ (see Fig. \ref{figset}). We impose $R=1.2$, $Q=(0.1\delta)^2$, or both constraints simultaneously for TOA and TDOA-based localization without priori knowledge of time reference mismatch (i.e., $J_{b_i}=0$). As shown in Fig. \ref{fig2}, as MS 2 moves apart from MS 1, the transmit power expenditure decreases when only the data rate constraint is imposed, due to the enhanced capability of beamforming to reduce the interference between the signals for the two MSs. In contrast, when only the localization constraint is imposed, the transmit power increases as the separation of the two MSs becomes larger. This is because, as seen in (\ref{eq:TOAEFIM}) and (\ref{eq:TDOAEFIMwb}), no part of the transmitted signal (\ref{eq:xp}) is to be treated as interference if the goal is localization. When imposing both rate and localization constraints, it is seen that the transmit power is larger than the worst-case power between both cases with only rate or localization constraints. 

\begin{figure}[t]
\begin{center}
\includegraphics[width=8cm]{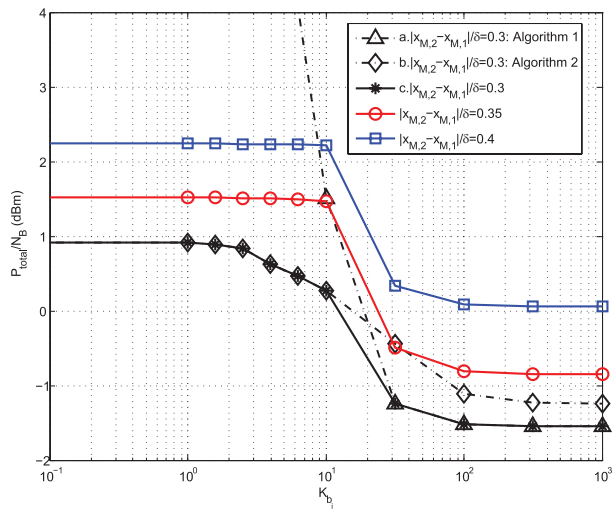}
\caption{Normalized per-BS transmit power with TDOA-based localization as a function of the prior information $J_{b_i} \propto K_{b_i}$ on time reference mismatch for $(M, N_B, N_M) = (4, 4, 2)$ with constraints $R=1.2$ and $Q=(0.1\delta)^2$ and the topology with BSs at the corners of the square area.} \label{fig6}
\end{center}
\end{figure}
From Fig. \ref{fig2}, it can also be observed that the minimum power in the presence of both rate and localization constraints when using TDOA-based localization does not converge to the minimum power with localization constraint only as MS $2$ moves away from MS $1$. This is unlike with TOA-based methods in which case does not affect the minimum power as the MSs become further apart. This can be explained since TDOA-based localization requires the BSs to transmit more power as it can be seen from the curves correspondingly to the localization constraint only. This increased power creates additional interference which must be properly managed in order to guarantee the rate constraint. As a result, the rate constraint affects the optimal beamforming design and the minimum power even when the MSs are apart.

In Fig. \ref{fig6}, we investigate the effect of the time reference mismatch on the minimum power for a network with $R=1.2$ and $Q=(0.1\delta)^2$, respectively, and as like the topology in Fig. $\ref{figset}$. For the case $|x_{M,2}-x_{M,1}|/\delta=0.3$, we show the performances of the first approach proposed in Sec. \ref{sec:TDOAbeam}, namely Algorithm $1$ with the LMI constraint (\ref{eq:TDOALMI}) in the lieu of (\ref{SchurTOA}), and of Algorithm $2$. It can be seen that for sufficiently large  $J_{b_i}$, the approach based on Algorithm $1$ with (\ref{eq:TDOALMI}) is to be preferred, while, if $J_{b_i}$ is smaller, Algorithm $2$ performs best. Also, we observe that the required transmit power decreases when the prior information $J_{b_i}$ increases. In accordance with Remark \ref{rem1}, for large $J_{b_i}$ it converges to that of TOA-based localization.
\begin{figure}[t]
\begin{center}
\includegraphics[width=8cm]{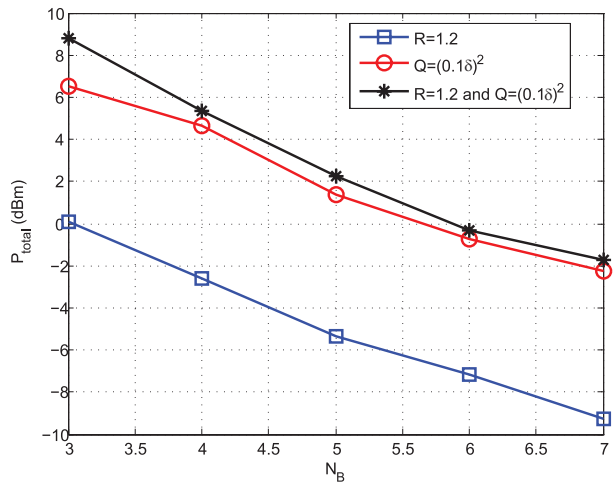}
\caption{Total transmit power with TOA-based localization as a function of the number of $N_B$ of BSs for $(M, N_M)=(4,2)$ and a random topology.} \label{fig3}
\end{center}
\end{figure}

\begin{figure}[t]
\begin{center}
\vspace{0.15cm}
\includegraphics[width=8cm]{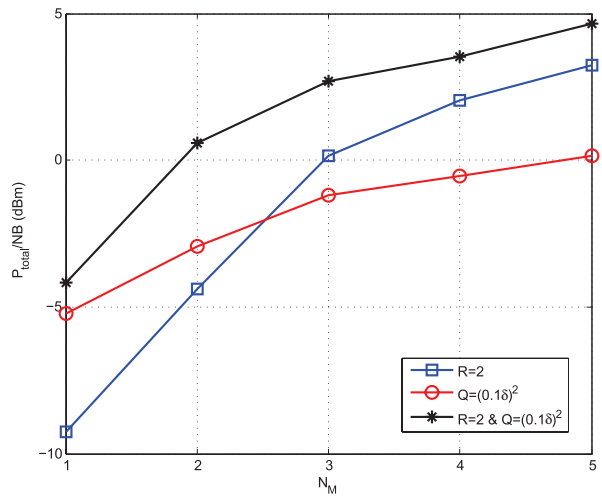}
\caption{Normalized per-BS transmit power with TOA-based localization as a function of the number $N_M$ of MSs for $(M,N_B)=(5,4)$ and a random topology.} \label{fig4}
\end{center}
\end{figure}
We next consider a set-up where both BSs and MSs are randomly and uniformly distributed in the square area of Fig. \ref{fig:sys}. Fig. \ref{fig3} and Fig. \ref{fig4} show the total power and the normalized per-BS power as a function of the number $N_B$ of BSs and the number $N_M$ of MSs, respectively. In Fig. \ref{fig3}, a network with $(M,N_M)=(4,2)$ is considered under $R=1.2$, $Q=(0.1\delta)^2$, or both constraints. It is observed that the total average transmit power decreases with the number of BSs, since more BSs increase the degrees of freedom available for optimization. Fig. \ref{fig4} considers a network for $(M,N_B)=(5,4)$ under $R=2$, $Q=(0.1\delta)^2$, or both constraints. The normalized average transmit power is shown to increase with the number $N_M$ of MSs. This is easily explained since, as $N_M$ increases, satisfying the rate constraints becomes more demanding in terms of power due to the increasingly more complex interference management task. Also, since more localization constraints are imposed for the additional MSs, the more power is required. The same quantitative behavior is observed for TDOA-based localization (not reported here).
\subsection{Robust Beamforming Design}\label{sec:simulrobust}
\begin{figure}[t]
\begin{center}
\includegraphics[width=8cm]{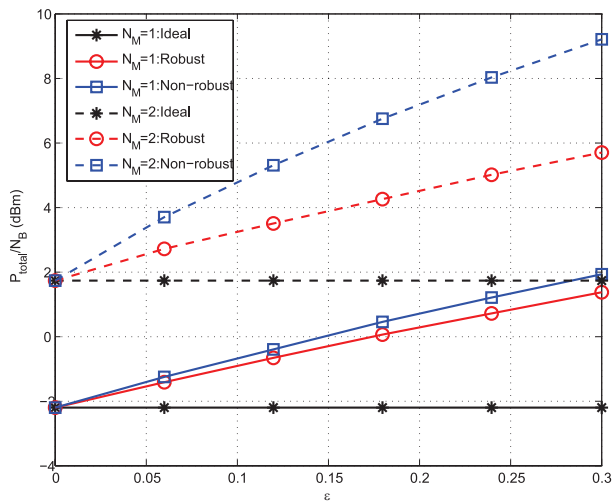}
\vspace{-0.2cm}
\caption{Normalized per-BS transmit power with TOA-based localization as a function of the uncertainty set size $\epsilon$ for $(M,N_B)=(4,4)$ and constraints $R=2$ and $Q=(0.1\delta)^2$.} \label{fig5}
\end{center}
\end{figure}
In this section, we evaluate the performance of the proposed robust transmission strategy for frequency-flat channels with TOA-based localization proposed in Sec. \ref{sec:TOArobust}.
In Fig. \ref{fig5}, we investigate the impact of imperfect network parameters with $R=2$ and $Q=(0.1\delta)^2$. We have $N_B=4$ BSs, each of which has $M=4$ transmit antennas and which are placed at the corners of the square region of Fig. \ref{fig:sys}. The nominal positions of the MSs are randomly and uniformly distributed in the square area. Moreover, for each nominal position $(\hat{d}_{ji}, \hat{\phi}_{ji})$ of a MS $i$, we uniformly generate pairs of the actual parameters $(d_{ji}, \phi_{ji})$ whose components lie in the uncertainty sets $S_{ji}^d$ and $S_{ji}^\phi$ in (\ref{intervalpara}), respectively. We set $\epsilon_{ji}^d=\epsilon\delta/2$ and $\epsilon_{ji}^\phi=2\epsilon$ (degrees) in (\ref{intervalpara}), where $\epsilon$ is a parameter defining the size of the uncertainty sets. For the CSI, we define the correlation matrix $\pmb{R}_{ji}=E[\pmb{s}_{ji}(\phi_{ji})\pmb{s}_{ji}^*(\phi_{ji})]$ where $\phi_{ji}$ is uniformly distributed in the uncertainty set. This matrix accounts for the average correlation of MSs. We consider the performance of the ideal scheme (labeled as ``ideal") that is designed based on the known parameters $(d_{ji}, \phi_{ji}, \pmb{h}_{ji})$ as described in Sec. \ref{sec:TOAbeam} and of the robust scheme introduced in Sec. \ref{sec:TOArobust}. For reference, we also consider a ``non-robust" scheme that uses the approach discussed in Sec. \ref{sec:TOAbeam} using the nominal parameters $(\hat{d}_{ji}, \hat{\phi}_{ji}, \hat{\pmb{h}}_{ji}=\pmb{s}_{ji}(\hat{\phi}_{ji}))$ as the actual parameters. For the latter case, we evaluate the required transmit power by solving problem (\ref{opttoa}) assuming the nominal parameters and then finding the minimal common power scaling of the so-obtained beamformers that guarantees the satisfaction of the rate and localization constraints. 
 
Fig. \ref{fig5} shows the required powers as a function of $\epsilon$. Clearly, a larger $\epsilon$ implies a larger required power for the non-robust and for the robust schemes due to the larger uncertainty sets. Moreover, the proposed robust strategy is seen to be effective in significantly reducing the power with respect to the non-robust approach. The effect is especially marked for a larger number of MSs given the need for a  more accurate design of the beamformers to handle interference. It is also observed that the performance with the robust strategy of TDOA-based localization introduced in Sec. \ref{sec:TDOArobust} has the same trend of TOA-based localization according to $\epsilon$ (not shown here due to space limits).
\subsection{Frequency-Selective Fading Channels}\label{sec:simulselect}
\begin{figure}[t]
\begin{center}
\includegraphics[width=8cm]{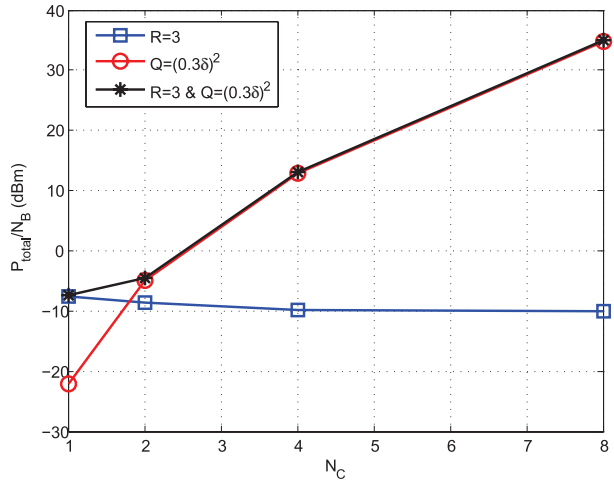}
\vspace{0.2cm}
\caption{Normalized per-BS transmit power with TOA-based localization for frequency-selective channels as a function of the number of blocks $N_C$ for $(M,N_B, N_M)=(4,4,2)$ with $N=32$ and constraints $R=3$ and $Q=(0.3\delta)^2$.} \label{fig7}
\end{center}
\end{figure}
For frequency-selective channels, we assume that all BS-MS pairs have three multi-paths, i.e., $L_i=3$. Moreover, we model the each $l$th path between BS $j$ and MS $i$ as $\pmb{h}_{ji,l}=|h_{ji,l}|\pmb{s}_{ji}(\phi_{ji}+\epsilon_{ji,l}^\phi)$, where $h_{ji,l}$ for $l=\{0,1,2\}$ are complex-valued zero-mean Gaussian random variables with exponential power decay and $\epsilon_{ji,l}^\phi$ is a random angle uniformly distributed over the interval $[-10,10]$ (degrees). 
\begin{table*}[t!]
\begin{center}
\caption{\lowercase{(a) LTE-based system parameters (b) Normalized per-BS transmit power with TOA-based localization}} \label{table1}
\includegraphics[width=18cm]{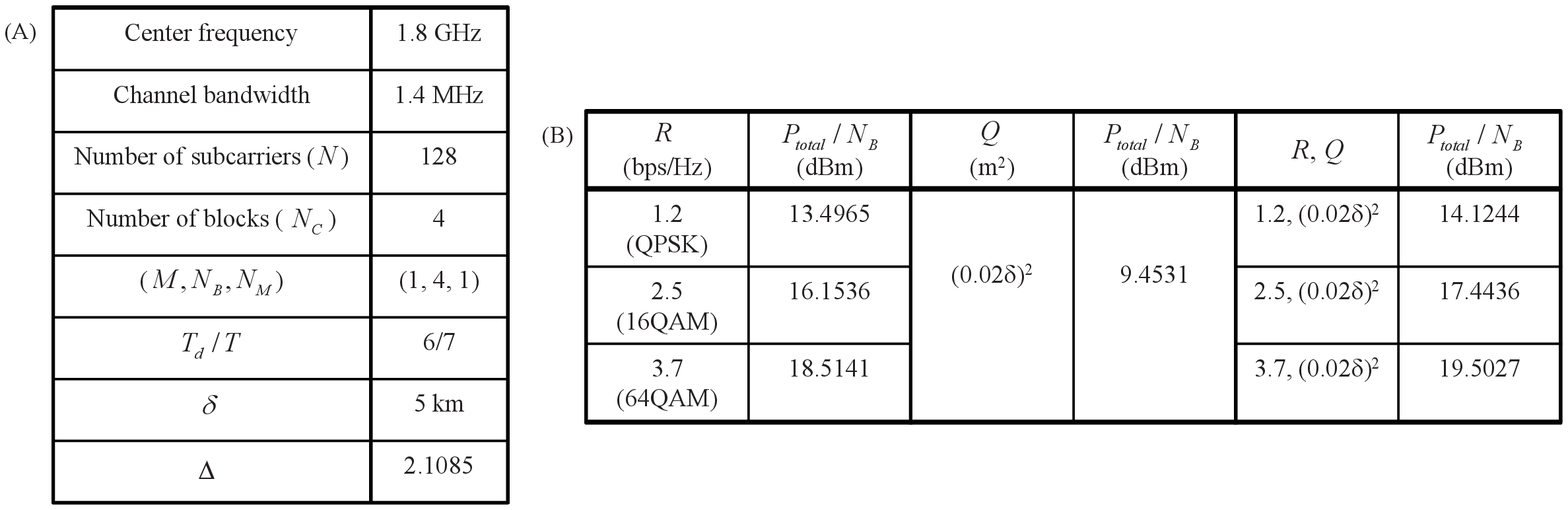}
\end{center}
\end{table*}
Fig. \ref{fig7} shows the required transmit power as a function of the number of frequency blocks $N_C$. We set $T_s=5\mu$s and consider an OFDM system with $N=32$ and $(M, N_B, N_M) = (4, 4, 2)$, where $N_B=4$ BSs are placed at the vertices of the square region of Fig. \ref{fig:sys}, while MS 1 is located in the center of square area, and MS $2$ is on the x-axis $50$m space away from MS $1$. We impose the constraints $R=3$ and $Q=(0.3\delta)^2$. As it can be seen, as $N_C$ increases, when considering only the data rate constraint, the transmit power expenditure decreases. This is thanks to the enhanced capability of beamforming to manage the interference in each smaller block. Conversely, when only the localization constraint is imposed, the transmit power increases with $N_C$. This is because, as seen, localization becomes impossible when the number of blocks is not small enough due to the excessive number of unknown parameters affecting the received signal (see Remark \ref{rem2}). Finally, when imposing both constraints, it is observed that the normalized transmit power is larger than the worst-case power between both constraints similar to the discussion around Fig. \ref{fig2}. 
\subsection{Case Study: LTE System with Localization Constraints}\label{sec:simulLTE}
In order to gain additional insights into the effects of localization constraints on the design of wireless systems, we evaluate the minimum transmit power required with the LTE-based system parameters summarized in Table \ref{table1} \cite{STDLTE3GPP} and assuming E-911 requirements for localization accuracy \cite{STDE911}. We consider a network with $N_B=4$ BSs placed at the corners of the square region of Fig. \ref{fig:sys}, while MS 1 is located at the center of square area. For frequency-selective channels, we use the same channel model as in Sec. \ref{sec:simulselect} with $L_i=3$. The reference distance $\Delta$ is chosen so that the path loss at distance $5$ km is $\zeta^2=-135$dB and we assume noise level $N_0=-112.5$dBm. The data rate requirements correspond to the peak downlink data rates for LTE with single-antenna receiver \cite{STDLTE3GPP}. Table \ref{table1}-(B) shows the minimum powers when imposing only the rate constraint $R$, only the E-911-based localization requirement of $Q=(0.02\delta)^2$, and both constraints. It is seen that the normalized transmit powers are slightly increased by imposing the localization accuracy constraints in addition to the data rate requirements.  
\section{Conclusions}\label{sec:con} 
In this paper, we investigated the beamforming design for location-aware distributed antenna systems under data communication and localization accuracy constraints. A number of iterative optimization algorithms were proposed that apply to frequency-flat and frequency-selective channels with TOA and TDOA-based localization measurements, and that operate in the presence of possibly imperfect knowledge of system parameters, such as CSI. The algorithms are based on rank-$1$ relaxation and DC programming. Moreover, for frequency-selective channels, we proposed a novel OFDM-based transmission strategy that provides a trade-off between rate and localization accuracy via grouping of the subcarriers. Extensive numerical results illustrate the interplay of the constraints on rate and localization accuracy. Among interesting open issues for future work, we point to the development of effective global optimization algorithms to tackle directly the non-convex beamforming design problems formulated in this work, which were handled here via efficient suboptimal strategies.                            
\newpage
\appendices
\section{Calculation of EFIMs for Frequency-Flat Fading Channels}\label{app:flatspeb}
In this Appendix, we calculate the EFIM (\ref{eq:TOAEFIM}) and (\ref{eq:TDOAEFIMwb}), respectively. The derivation follows the same main steps as in \cite{Shen10TIT} with the main differences that \cite{Shen10TIT} focuses on wide-band impulsive signals over multi-path real channels, while here we focus on the stream signal (\ref{eq:xp}) over flat-fading complex channels. For simplicity, we assume that the real training sequences are employed. 
\subsection{Calculation of EFIM (\ref{eq:TOAEFIM}) for TOA-based Localization}\label{app:flatTOA}
First of all, for TOA-based localization, i.e., $b_i=0$, we define the unknown parameter vector for MS $i$ as $\pmb{\theta}_i=[\pmb{p}_{M,i}^T \,\, \pmb{\alpha}_{1i}^T \,\, \cdots \,\,\pmb{\alpha}_{N_Bi}^T]^T$, where $\pmb{\alpha}_{ji}=[\text{Re}\{\alpha_{ ji}^{(1)}(\pmb{w}_{j1})\}\,\, \text{Im}\{\alpha_{ji}^{(1)}(\pmb{w}_{j1})\} \cdots \text{Re}\{\alpha_{ji}^{(N_M)}(\pmb{w}_{jN_M})\}\,\,$ $\text{Im}\{\alpha_{ji}^{(N_M)}(\pmb{w}_{jN_M})\}]^T$. Similar to \cite{Shen10TIT}, assuming that the MS $i$ is localizable, i.e., that $\pmb{p}_{M,i}$ can be determined by $\tau_{ji}$ for $j \in \mathcal{N}_B$ via triangulation, the mapping of $\pmb{\theta}_i$ to the parameter vector $\tilde{\pmb{\theta}}_i=[\tilde{\pmb{\theta}}_{1i}^T \,\, \tilde{\pmb{\theta}}_{2i}^T \,\, \cdots \,\,\tilde{\pmb{\theta}}_{N_Bi}^T]^T$, where $\tilde{\pmb{\theta}}_{ji}=[\tau_{ji} \,\, \pmb{\alpha}_{ji}^T]^T$ is a bijection. We denote the vector of all received pilot signals as $\pmb{y}_i=[\pmb{y}_{1i}^{(p)T}\cdots\pmb{y}_{N_Bi}^{(p)T}]^T$, hence by using the vector representation $\pmb{y}_{ji}^{(p)}$ for $y_{ji}^{(p)}(t)$ over the training phase of duration $T_p$. Then, we can express the FIM $\pmb{J}_{\pmb{\theta}_i}$ for the parameter vector $\pmb{\theta}_i$ as a function of the FIM $\pmb{J}_{\tilde{\pmb{\theta}}_i}$ for $\tilde{\pmb{\theta}}_i$ as $\pmb{J}_{\pmb{\theta}_i}=\pmb{T}_i\pmb{J}_{\tilde{\pmb{\theta}}_i}\pmb{T}_i^T$ \cite{Shen10TIT}, where the Jacobian matrix $\pmb{T}_i \in \mathrm{R}^{(2+2N_BN_M) \times N_B(2N_M+1)}$ for the transformation from $\pmb{\theta}_i$ to $\tilde{\pmb{\theta}}_i$ and the FIM $\pmb{J}_{\tilde{\pmb{\theta}}_i} \in \mathrm{R}^{N_B(2N_M+1) \times N_B(2N_M+1)}$ are given by
\begin{subequations}\label{eq:Tandjtilde}
\begin{eqnarray}
&&\hspace{0cm}\pmb{T}_i = \frac{\partial\tilde{\pmb{\theta}}_i}{\partial\pmb{\theta}_i}=\left[ \begin{array}{ccc}
\pmb{G}_{1i} & \cdots & \pmb{G}_{N_Bi}\\
\pmb{D}_1 & \cdots& \pmb{D}_{N_B}\end{array} \right]\\
&&\hspace{-1.6cm}\text{and}\nonumber\\
&&\hspace{-1.4cm}\left[\pmb{J}_{\tilde{\pmb{\theta}}_i}\right]_{n,m} = \text{E}_{\pmb{y}_i}\left\{\left[\frac{\partial\ln f(\pmb{y}_i|\tilde{\pmb{\theta}}_i)}{\partial\tilde{\pmb{\theta}}_{i,n}}\right]\left[\frac{\partial\ln f(\pmb{y}_i|\tilde{\pmb{\theta}}_i)}{\partial\tilde{\pmb{\theta}}_{i,m}}\right]^*\right\}
\end{eqnarray}
\end{subequations}
where $\pmb{G}_{ji} \in \mathrm{R}^{2 \times (2N_M+1)}
=\frac{1}{c}[\pmb{q}_{ji} \,\, \pmb{0}_{2 \times 2N_M}]$ and 
$\pmb{D}_j \in \mathrm{R}^{2N_MN_B\times (2N_M+1)}$ has all zero elements except for $[\pmb{D}_j]_{(2N_M(j-1)+1:2N_Mj,\,2:2N_M+1)}=\pmb{I}_{2N_M}$. Moreover, $\pmb{J}_{\tilde{\pmb{\theta}}_i}=\text{diag}\left\{\pmb{\Psi}_{1i},\pmb{\Psi}_{2i},\dots,\pmb{\Psi}_{N_Bi}\right\}$, where
\begin{eqnarray}\label{eq:psi}
&&\hspace{-1.2cm}\pmb{\Psi}_{ji} \in \mathrm{R}^{(2N_M+1)\times (2N_M+1)}\nonumber\\
&&\hspace{-0.6cm}= \text{E}_{\pmb{y}_i}\left\{\left[\frac{\partial\ln f(\pmb{y}_i|\tilde{\pmb{\theta}}_i)}{\partial\tilde{\pmb{\theta}}_{ji}}\right]\left[\frac{\partial\ln f(\pmb{y}_i|\tilde{\pmb{\theta}}_i)}{\partial\tilde{\pmb{\theta}}_{ji}}\right]^*\right\}.
\end{eqnarray}
By using Slepian-Bang's formula \cite[Sec. 3.9]{Kay93Book}, the matrix ($\ref{eq:psi}$) is evaluated as in $\pmb{\Psi}_{ji}=[
8\pi^2n_p\beta^2\text{SNR}_{ji}(\pmb{W}_j)$ $\,\,\pmb{0}_{2N_M}^T; \pmb{0}_{2N_M}\,\,\frac{2n_p}{N_0}\pmb{I}_{2N_M}]$. Moreover, using the relationship of $\pmb{J}_{\pmb{\theta}_i}=\pmb{T}_i\pmb{J}_{\tilde{\pmb{\theta}}_i}\pmb{T}_i^T$ we can write 
\begin{equation}\label{eq:reFIM}
\pmb{J}_{\pmb{\theta}_i}=\left[\begin{array}{cc}
\pmb{A} & \pmb{B}\\
\pmb{B}^T & \pmb{C}\end{array} \right],
\end{equation}
where $\pmb{A}=\sum_{j \in \mathcal{N}_B} \pmb{G}_{ji}\pmb{\Psi}_{ji}\pmb{G}_{ji}^T=\frac{8\pi^2n_p\beta^2}{c^2} \sum_{j \in \mathcal{N}_B}\text{SNR}_{ji}(\pmb{W}_j)\pmb{q}_{ji}\pmb{q}_{ji}^T$, $\pmb{B}=[\pmb{G}_{1i}\pmb{\Psi}_{1i}\pmb{D}^T \,\, \cdots \,\, \pmb{G}_{N_Bi}\pmb{\Psi}_{N_Bi}\pmb{D}^T]=\pmb{0}_{2 \times 2N_BN_M}$ and $\pmb{C}=\text{diag}\left\{\pmb{D}\pmb{\Psi}_{1i}\pmb{D}^T,\cdots, \pmb{D}\pmb{\Psi}_{N_Bi}\pmb{D}^T\right\}=\frac{2n_p}{N_0}\pmb{I}_{2N_BN_M}$, where $\pmb{D}=[\pmb{0}_{2N_M}\,\,\pmb{I}_{2N_M}]$. Applying the Schur complement, $\pmb{J}_{i,\text{TOA}}(\pmb{W})=\pmb{A}-\pmb{B}\pmb{C}^{-1}\pmb{B}^T$ (see, e.g., \cite{Zhang05Book}), we can finally obtain the EFIM $\pmb{J}_{i,\text{TOA}}(\pmb{W})$ as (\ref{eq:TOAEFIM}).
\subsection{Calculation of EFIM (\ref{eq:TDOAEFIMwb}) for TDOA-based Localization}\label{app:flatTDOA}
The derivation of EFIM (\ref{eq:TDOAEFIMwb}) for TDOA-based localization is similar and the main difference here is the need to consider the time reference mismatch $b_i$ between the BSs and MS $i$ in (\ref{eq:rxsig}). We define the unknown parameter vector for MS $i$ with the addition of the mismatch $b_i$ as
$\pmb{\theta}_i=[\pmb{p}_{M,i}^T \,\, b_i \,\, \pmb{\alpha}_{1i}^T \,\, \cdots \,\,\pmb{\alpha}_{N_Bi}^T]^T$ and the mapping $\pmb{\theta}_i$ to $\tilde{\pmb{\theta}}_i$ is a bijection assuming that the MS $i$ is localizable as stated above. Note that the quantity $\tau_{ji}$ in (\ref{tau}) represents the effective delay, including the timing offset, between BS $j \in \mathcal{N}_B$ and MS $i \in \mathcal{N}_M$. Then, we have the relationship of $\pmb{J}_{\pmb{\theta}_i}=\pmb{T}_i\pmb{J}_{\tilde{\pmb{\theta}}_i}\pmb{T}_i^T+\pmb{J}_{p_i}$, where $\pmb{T}_i \in \mathrm{R}^{(3+2N_BN_M) \times N_B(2N_M+1)}$ and $\pmb{J}_{\tilde{\pmb{\theta}}_i} \in \mathrm{R}^{N_B(2N_M+1) \times N_B(2N_M+1)}$ are given by
\begin{subequations}\label{TDOA:t}
\begin{eqnarray}
&&\hspace{-1cm}\pmb{T}_i = \frac{\partial\tilde{\pmb{\theta}}_i}{\partial\pmb{\theta}_i}=\left[ \begin{array}{ccc}
\pmb{G}_{1i} & \cdots &\pmb{G}_{N_Bi}\\
\pmb{g}^T & \cdots & \pmb{g}^T \\ 
\pmb{D}_1 &\cdots& \pmb{D}_{N_B} \end{array} \right]\\
&& \hspace{-2.6cm}\text{and}\nonumber\\
&&\hspace{-1cm}\pmb{J}_{\tilde{\pmb{\theta}}_i}=\text{diag}\left\{\pmb{\Psi}_{1i},\pmb{\Psi}_{2i},\dots,\pmb{\Psi}_{N_Bi}\right\},
\end{eqnarray}
\end{subequations}
where the block matrices $\pmb{G}_{ji} \in \mathrm{R}^{2 \times (2N_M+1)}$, $\pmb{D}_j \in \mathrm{R}^{2N_MN_B\times (2N_M+1)}$ and $\pmb{\Psi}_{ji} \in \mathrm{R}^{(2N_M+1)\times (2N_M+1)}$ are the same given as (\ref{eq:Tandjtilde}) and (\ref{eq:psi}), respectively and $\pmb{g}=[1 \,\,\pmb{0}_{2N_M}^T]^T$. The joint pdf of observation and parameters can be expressed as $f(\pmb{y}_i,\pmb{\theta}_i)=f(\pmb{y}_i|\pmb{\theta}_i)+f(\pmb{\theta}_i)$, and thus the FIM becomes $\pmb{J}_{\pmb{\theta}_i}=\pmb{J}_{w_i}+\pmb{J}_{p_i}$, where $\pmb{J}_{w_i}=\text{E}_{\pmb{y}_i,\pmb{\theta}_i}\left\{\left[\frac{\partial\ln f(\pmb{y}_i|\pmb{\theta}_i)}{\partial\pmb{\theta}_i}\right]\left[\frac{\partial\ln f(\pmb{y}_i|\pmb{\theta}_i)}{\partial\pmb{\theta}_i}\right]^*\right\}$ and $\pmb{J}_{p_i}=\text{E}_{\pmb{\theta}_i}\left\{\left[\frac{\partial\ln f(\pmb{\theta}_i)}{\partial\pmb{\theta}_i}\right]\left[\frac{\partial\ln f(\pmb{\theta}_i)}{\partial\pmb{\theta}_i}\right]^*\right\}$ are the FIMs from the observations and the priori knowledge, respectively. $\pmb{J}_{w_i}$ can be calculated by taking the expectation of $\pmb{J}_{\pmb{\theta}_i}$ over the parameter vector $\pmb{\theta}_i$, and $\pmb{J}_{p_i} \in \mathrm{R}^{(3+2N_BN_M) \times (3+2N_BN_M)}$ has all zero components except $[\pmb{J}_{p_i}]_{3,3}=J_{b_i}$ since the MS $i$'s position and effective channel parameters are deterministic, but $b_i$ is random. Therefore, by using Slepian-Bang's formula \cite[Sec. 3.9]{Kay93Book}, the FIM $\pmb{J}_{\pmb{\theta}_i}$ can be again written as in (\ref{eq:reFIM}), where 
$\pmb{A}=\sum_{j \in \mathcal{N}_B} \pmb{G}_{ji}\pmb{\Psi}_{ji}\pmb{G}_{ji}^T=\frac{8\pi^2n_p\beta^2}{c^2} \sum_{j\in \mathcal{N}_B}\text{SNR}_{ji}(\pmb{W}_j)\pmb{q}_{ji}\pmb{q}_{ji}^T$, $\pmb{B}=[\sum_{j\in \mathcal{N}_B}\pmb{G}_{ji}\pmb{\Psi}_{ji}\pmb{g}\,\,\,\,\pmb{G}_{1i}\pmb{\Psi}_{1i}\pmb{D}^T \,\, \cdots \,\,\,\, \pmb{G}_{N_Bi}\pmb{\Psi}_{N_Bi}\pmb{D}^T]=[\frac{8\pi^2n_p\beta^2}{c}\sum_{j\in \mathcal{N}_B}\text{SNR}_{ji}$ $(\pmb{W}_j)\pmb{q}_{ji}\,\,\,\,\,\,\pmb{0}_{2 \times 2N_BN_M}]$ and $\pmb{C}=\text{diag}\{\sum_{j\in \mathcal{N}_B}\pmb{g}^T\pmb{\Psi}_{ji}\pmb{g}+J_{b_i},\pmb{D}\pmb{\Psi}_{1i}\pmb{D}^T,\dots, \pmb{D}\pmb{\Psi}_{N_Bi}\pmb{D}^T\}
=\text{diag}\{8\pi^2n_p\beta^2$ $\sum_{j\in \mathcal{N}_B}\text{SNR}_{ji}$ $(\pmb{W}_j)+J_{b_i}\,\,$ $\frac{2n_p}{N_0}\pmb{I}_{2N_BN_M}\}$, where $\pmb{D}=[\pmb{0}_{2N_M}\,\,\pmb{I}_{2N_M}]$. By applying the Schur complement (see, e.g., \cite{Zhang05Book}), we can finally obtain the EFIM of TDOA as (\ref{eq:TDOAEFIMwb}).
\section{Derivation and 
Properties of Algorithm $1$}\label{app:TOAbeam}
Recall that we are interested in solving problem (\ref{opttoa}). By utilizing the positive semi-definiteness of EFIM (\ref{eq:TOAEFIM}), the localization constraint (\ref{eq:toaspebreq}) can be reformulated as a (convex) LMI constraint (see e.g., \cite{Shen12Globecom, Li13Arxiv}) by introducing the auxiliary matrix $\pmb{M}_i \succeq 0$ and applying the Schur complement condition for positive semi-definiteness (see, e.g., \cite{Zhang05Book}) as follows:   
\begin{subequations}\label{toasdp}
\begin{eqnarray}
&& \hspace{-1.7cm}{  \left( \hspace{-0.2cm}\begin{array}{cc}
\pmb{M}_i & \pmb{I}  \\
\hspace{-0.3cm}\pmb{I} & \hspace{-0.2cm}\frac{8\pi^2n_p\beta^2}{c^2N_0}\sum_{j \in \mathcal{N}_B}\sum_{k \in \mathcal{N}_M}\xi_{ji}^{(k)}(\pmb{\Sigma}_{jk})\pmb{J}_\phi(\phi_{ji}) \hspace{-0.2cm} \end{array} \right) \hspace{-0.1cm}\succeq\hspace{-0.1cm} 0,}\\
&& \hspace{1.2cm} \text{tr}\left\{\pmb{M}_i\right\} \le Q_i,\\
&& \hspace{1.2cm} {\pmb{M}_i \succeq 0,} \hspace{0.5cm} \forall i \in \mathcal{N}_M.
\end{eqnarray}
\end{subequations}
Now, problem (\ref{eq:min})-(\ref{eq:psd}), with (\ref{toasdp}) in lieu of (\ref{eq:toaspebreq}), must be solved over both $\pmb{\Sigma}_{ji}$ and $\pmb{M}_i$. 

As mentioned, the resulting problem is convex except for the constraints 
(\ref{eq:toaratereq}) and (\ref{eq:rank}). For the former, we observe that its left-hand side can be written as a DC functions. Therefore the constraint (\ref{eq:toaratereq}) can be handled by invoking the MM algorithm \cite{Beck10Book}, which solves a sequence of convex problems obtained by linearizing the non-convex part of the constraint. It is known that the MM algorithm converges to a stationary point of the optimization problem \cite{Hunter04Amer}. Also, each iteration of MM algorithm provides a feasible solution of the relaxed problem since the rate used in the constraint (\ref{eq:toaratereq}) is always a lower bound of the actual rate due to the concavity of the log function. 

It hence remains to discuss the rank constraint (\ref{eq:rank}). Here, following the approach commonly used in related problems, we relax this constraint. Having obtained a stationary point $\pmb{\Sigma}_{ji}^{\rm{opt}}$ of the relaxed problem via the MM algorithm, we obtain the beamforming vector $\hat{\pmb{w}}_{ji}$ by the principal eigenvector approximation of the covariance matrix obtained as $\hat{\pmb{w}}_{ji}=\sqrt{\lambda_{\rm{max}}(\pmb{\Sigma}_{ji}^{\rm{opt}})}\pmb{v}_{\rm{max}}(\pmb{\Sigma}_{ji}^{\rm{opt}})$, $\forall j \in \mathcal{N}_B$ and $\forall i \in \mathcal{N}_M$ (see, e.g., \cite{Luo10SPM, Beng99Allerton}). If $\hat{\pmb{w}}_{ji}$ is feasible, we finally decide the optimal beamforming vector as $\pmb{w}_{ji}^{\rm{opt}}=\hat{\pmb{w}}_{ji}$. Otherwise, we rescale $\hat{\pmb{w}}_{ji} \leftarrow (1+\delta_{\rm{inc}})\hat{\pmb{w}}_{ji}$ for a positive integer $\delta_{\rm{inc}}$ until the $\hat{\pmb{w}}_{ji}$ is feasible. Note that, due to the monotonicity of the rate (\ref{eq:toaratereq}) and EFIM (\ref{eq:toaspebreq}) with respect to $\delta_{\rm{inc}}$, it is always possible to find a scaling factor $\delta_{\rm{inc}}$ such that $\pmb{w}_{ji}^{\rm{opt}}$ provides a feasible solution to the original problem (\ref{opttoa}).
\section{Proof of Lemma \ref{lem1}}\label{app:QfiTDOA}
The rightmost inequality in (\ref{QfiTDOArel}) is satisfied by the fact that $\zeta_{ji}^*=\zeta_{ji}^{\rm{L}}$ is the solution of problem (\ref{eq:Ropt}). 

For the leftmost inequality in (\ref{QfiTDOArel}), we need to prove that $\pmb{J}_\phi(\phi_{ji}, \phi_{j'i})-\pmb{Q}_{\phi}(\hat{\phi}_{ji}, \hat{\phi}_{j'i})$ is positive semidefinite for all $\phi_{ji} \in S_{ji}^\phi$, $\phi_{j'i} \in S_{j'i}^\phi$ and $1 \le j <j' \le N_B$. To this end, we calculate
\begin{eqnarray}
&&\hspace{-0.75cm}\left[\hspace{-0.05cm}\pmb{J}_\phi(\phi_{ji}\hspace{-0.05cm}, \phi_{j'i})\hspace{-0.1cm}-\hspace{-0.06cm}\pmb{Q}_{\phi}(\hat{\phi}_{ji}\hspace{-0.05cm}, \hat{\phi}_{j'i})\hspace{-0.05cm}\right]_{1,1}=\delta_{(j,j')i}-\sin\phi_{ji}^{+}\sin\phi_{ji}^{-}-\sin\phi_{j'i}^{+}\sin\phi_{j'i}^{-}+A+B\nonumber\\
&&\hspace{-0.75cm}\left[\hspace{-0.05cm}\pmb{J}_\phi(\phi_{ji}\hspace{-0.05cm}, \phi_{j'i})\hspace{-0.1cm}-\hspace{-0.06cm}\pmb{Q}_{\phi}(\hat{\phi}_{ji}\hspace{-0.05cm}, \hat{\phi}_{j'i})\hspace{-0.05cm}\right]_{1,2}=\left[\hspace{-0.05cm}\pmb{J}_\phi(\phi_{ji}\hspace{-0.05cm}, \phi_{j'i})\hspace{-0.1cm}-\hspace{-0.06cm}\pmb{Q}_{\phi}(\hat{\phi}_{ji}\hspace{-0.05cm}, \hat{\phi}_{j'i})\hspace{-0.05cm}\right]_{2,1}=\cos\phi_{ji}^{+}\sin\phi_{ji}^{-}+\cos\phi_{j'i}^{+}\sin\phi_{j'i}^{-}+C\nonumber\\
&&\hspace{-0.75cm}\left[\hspace{-0.05cm}\pmb{J}_\phi(\phi_{ji}\hspace{-0.05cm}, \phi_{j'i})\hspace{-0.1cm}-\hspace{-0.06cm}\pmb{Q}_{\phi}(\hat{\phi}_{ji}\hspace{-0.05cm}, \hat{\phi}_{j'i})\hspace{-0.05cm}\right]_{2,2}=\delta_{(j,j')i}+\sin\phi_{ji}^{+}\sin\phi_{ji}^{-}+\sin\phi_{j'i}^{+}\sin\phi_{j'i}^{-}-A+B
\end{eqnarray} 
where $\phi_{ji}^+=\phi_{ji}+\hat{\phi}_{ji}$, $\phi_{ji}^-=\phi_{ji}-\hat{\phi}_{ji}$, $\delta_{(j,j')i}=\sin\epsilon_{ji}^\phi+\sin\epsilon_{j'i}^\phi+4\sin\frac{\epsilon_{ji}^\phi+\epsilon_{j'i}^\phi}{2}$, $A=\cos(\hat{\phi}_{ji}+\hat{\phi}_{j'i})-\cos(\phi_{ji}+\phi_{j'i})$, $B=\cos(\hat{\phi}_{ji}-\hat{\phi}_{j'i})-\cos(\phi_{ji}-\phi_{j'i})$ and $C=\sin(\hat{\phi}_{ji}+\hat{\phi}_{j'i})-\sin(\phi_{ji}+\phi_{j'i})$. 
Since $\pmb{J}_\phi(\phi_{ji}, \phi_{j'i})-\pmb{Q}_{\phi}(\hat{\phi}_{ji}, \hat{\phi}_{j'i}) \in R^{2\times2}$ is symmetric, we have the inequality $\pmb{J}_\phi(\phi_{ji}, \phi_{j'i})-\pmb{Q}_{\phi}(\hat{\phi}_{ji}, \hat{\phi}_{j'i}) \succeq 0$ if and only if $[\pmb{J}_\phi(\phi_{ji}, \phi_{j'i})-\pmb{Q}_{\phi}(\hat{\phi}_{ji}, \hat{\phi}_{j'i})]_{1,1} \ge 0$, $[\pmb{J}_\phi(\phi_{ji}, \phi_{j'i})-\pmb{Q}_{\phi}(\hat{\phi}_{ji}, \hat{\phi}_{j'i})]_{2,2} \ge 0$ and $|\pmb{J}_\phi(\phi_{ji}, \phi_{j'i})-\pmb{Q}_{\phi}(\hat{\phi}_{ji}, \hat{\phi}_{j'i})| \ge 0$ \cite{Boyd04Book}. The above three conditions can be summarized as $\delta_{(j,j')i} \ge \sin\phi_{ji}^{+}\sin\phi_{ji}^{-}+\sin\phi_{j'i}^{+}\sin\phi_{j'i}^{-}-A-B$, $\delta_{(j,j')i} \ge -\sin\phi_{ji}^{+}\sin\phi_{ji}^{-}-\sin\phi_{j'i}^{+}\sin\phi_{j'i}^{-}+A-B$ and $\delta_{(j,j')i} \ge \sqrt{D}-B$, respectively, where $D=\sin^2\phi_{ji}^{-}+\sin^2\phi_{j'i}^{-}+4\sin^2\frac{\phi_{ji}^{-}+\phi_{j'i}^{-}}{2}+2\sin\phi_{ji}^{-}\sin\phi_{j'i}^{-}\cos(\phi_{ji}^{+}-\phi_{j'i}^{+})-4\sin\frac{\phi_{ji}^{-}+\phi_{j'i}^{-}}{2}$ $\cos\frac{\phi_{ji}^{+}-\phi_{j'i}^{+}}{2}(\sin\phi_{ji}^{-}+\sin\phi_{j'i}^{-})$. Since $\left|\phi_{ji}^-\right| \le \epsilon_{ji}^\phi$, $\left|\phi_{j'i}^-\right| \le \epsilon_{j'i}^\phi$, $|A| \le 2\sin\frac{\epsilon_{ji}^\phi+\epsilon_{j'i}^\phi}{2}$ and $|B| \le 2\sin\frac{\epsilon_{ji}^\phi+\epsilon_{j'i}^\phi}{2}$, these conditions are always satisfied for all $\phi_{ji} \in S_{ji}^\phi$, $\phi_{j'i} \in S_{j'i}^\phi$. This concludes the proof.
\section{Calculation of the EFIM for Frequency-Selective Fading Channels}\label{app:ofdmTOA}
The derivation of the EFIM (\ref{FIMTOAselec}) is similar to the derivation of (\ref{eq:TOAEFIM}) discussed in Appendix \ref{app:flatspeb}. The parameter vector $\pmb{\theta}_i$ associated with the OFDM signal model of (\ref{ofdmpilot}) is represented as $\pmb{\theta}_i=[\pmb{p}_{M,i}^T \,\, \pmb{\alpha}_{1i,1}^T\cdots\pmb{\alpha}_{1i,N_C}^T\cdots \pmb{\alpha}_{N_Bi,1}^T\cdots$ $\pmb{\alpha}_{N_Bi,N_C}^T]^T$, where $\pmb{\alpha}_{ji,b}=[\text{Re}\{\pmb{\alpha}_{ji,b}^{(1)}(\pmb{w}_{j1,b})^T\} \,\,\text{Im}\{\pmb{\alpha}_{ji,b}^{(1)}(\pmb{w}_{j1,b})^T\}\dots\text{Re}\{\pmb{\alpha}_{ji,b}^{(N_M)}(\pmb{w}_{jN_M,b})^T\}$ $\,\,\text{Im}\{\pmb{\alpha}_{ji,b}^{(N_M)}(\pmb{w}_{jN_M,b})^T\}]^T$ for $b \in \mathcal{N}_C$. Assuming that the MS $i$ is localizable, the mapping between $\pmb{\theta}_i$ and $\tilde{\pmb{\theta}}_i=[\tilde{\pmb{\theta}}_{1i,1}^T \cdots \tilde{\pmb{\theta}}_{1i,N_C}^T \cdots$ $ \tilde{\pmb{\theta}}_{N_Bi,1}^T\cdots \tilde{\pmb{\theta}}_{N_Bi,N_C}^T]^T$, where we defined $\tilde{\pmb{\theta}}_{ji,b}=[\tau_{ji} \,\, \pmb{\alpha}_{ji,b}^T]^T$, is a bijection. Therefore, we have the relationship $\pmb{J}_{\pmb{\theta}_i}=\pmb{T}_i\pmb{J}_{\tilde{\pmb{\theta}}_i}\pmb{T}_i^T$, where $\pmb{T}_i$ is the $(2+2N_BN_MN_CL_i)\times (N_BN_C(1+2N_ML_i))$  Jacobian matrix for the transformation from $\pmb{\theta}_i$ to $\tilde{\pmb{\theta}}_i$ and $\pmb{J}_{\tilde{\pmb{\theta}}_i}$ is the $N_BN_C(1+2N_ML_i) \times N_BN_C(1+2N_ML_i)$ matrix. Denote as $\pmb{y}_i$ the overall received signal by the MS $i$ from all the BSs, namely $\pmb{y}_i=[\pmb{y}_{1i,1}^{(p)T}\cdots\pmb{y}_{1i,N_C}^{(p)T}\cdots$ $\pmb{y}_{N_Bi,1}^{(p)T} \cdots\pmb{y}_{N_Bi,N_C}^{(p)T}]^T$. We can now calculate the matrices
\begin{eqnarray}\label{TDOA:tg}
&& \hspace{-0.7cm} \pmb{T}_i = \frac{\partial\tilde{\pmb{\theta}}_i}{\partial\pmb{\theta}_i}=\left[ \hspace{-0.2cm}\begin{array}{ccccccc}
\pmb{G}_{1i} & \hspace{-0.3cm}\cdots & \hspace{-0.2cm}\pmb{G}_{1i} & \hspace{-0.2cm}\cdots & \hspace{-0.2cm}\pmb{G}_{N_Bi} &\hspace{-0.2cm} \cdots &\hspace{-0.2cm} \pmb{G}_{N_Bi}\\
\pmb{D}_{1,1} & \hspace{-0.3cm}\cdots& \hspace{-0.2cm}\pmb{D}_{1,N_C} &\hspace{-0.2cm}\cdots& \hspace{-0.2cm}\pmb{D}_{N_B,1} &\hspace{-0.2cm}\cdots&\hspace{-0.2cm}\pmb{D}_{N_B,N_C} \end{array} \right]
\end{eqnarray}
and
\begin{equation}
\pmb{J}_{\tilde{\pmb{\theta}}_i}=\text{diag}\left\{\pmb{\Psi}_{1i,1},\dots, \pmb{\Psi}_{1i,N_C},\dots,\pmb{\Psi}_{N_Bi,1},\dots,\pmb{\Psi}_{N_Bi,N_C}\right\},
\end{equation}
where $\pmb{G}_{ji}=\frac{1}{c}[\pmb{q}_{ji}\,\,\pmb{0}_{2\times2N_ML_i}]$ and $\pmb{D}_{j,b} \in \mathrm{R}^{2N_BN_MN_CL_i\times (2N_ML_i+1)}$ has all zero elements except for
$[\hspace{-0.02cm}\pmb{D}_{j\hspace{-0.05cm},b}\hspace{-0.02cm}]_{(\hspace{-0.03cm}2N_ML_iN_C\hspace{-0.03cm}(\hspace{-0.03cm}j-1\hspace{-0.03cm})\hspace{-0.03cm}+\hspace{-0.02cm}2N_ML_i\hspace{-0.03cm}(\hspace{-0.03cm}b-1\hspace{-0.03cm})\hspace{-0.03cm}+\hspace{-0.02cm}1:2N_ML_i\hspace{-0.03cm}(\hspace{-0.03cm}(\hspace{-0.03cm}j-1\hspace{-0.03cm})N_C\hspace{-0.03cm}+\hspace{-0.02cm}b\hspace{-0.03cm})\hspace{-0.03cm},\hspace{-0.03cm}2:2N_ML_i\hspace{-0.03cm}+\hspace{-0.02cm}1\hspace{-0.03cm})}$ $=\pmb{I}_{2N_ML_i}$. By using Slepian-Bang's formula \cite[Sec. 3.9]{Kay93Book}, the matrix $\pmb{\Psi}_{ji,b} \in \mathrm{R}^{(2N_ML_i+1)\times (2N_ML_i+1)}$ can be calculated as 
\begin{equation}
\pmb{\Psi}_{ji,b}=\frac{2}{N_0}\left[\begin{array}{cc}\pmb{J}_{\tau_{ji,b}}&\pmb{J}_{ji,b}^T\\
\pmb{J}_{ji,b}&\pmb{J}_{\alpha_{ji,b}}\end{array}\right], 
\end{equation}
where $\pmb{J}_{\tau_{ji,b}}=\frac{4\pi^2}{T_s^2}\sum_{k \in \mathcal{N}_M}\pmb{\alpha}_{ji,b}^{(k)*}(\pmb{w}_{jk,b})\pmb{F}_{L_i,b}^*\pmb{K}_b^2\pmb{F}_{L_i,b}$ $\pmb{\alpha}_{ji,b}^{(k)}(\pmb{w}_{jk,b})$, $\pmb{J}_{ji,b}=\frac{2\pi}{T_s}[\pmb{J}_{ji,b}^{(1)T} \dots \pmb{J}_{ji,b}^{(N_M)T}]^T$, with $\pmb{J}_{ji,b}^{(k)}=[
\text{Im}\{\pmb{F}_{L_i,b}^*\pmb{K}_b\pmb{F}_{L_i,b}\pmb{\alpha}_{ji,b}^{(k)}$ $(\pmb{w}_{jk,b})\};
-\text{Re}\{\pmb{F}_{L_i,b}^*\pmb{K}_b\pmb{F}_{L_i,b}\pmb{\alpha}_{ji,b}^{(k)}(\pmb{w}_{jk,b})\}]$,  $\pmb{J}_{\alpha_{ji,b}}=\text{diag}\{\pmb{J}_{\alpha_{ji,b}}^{(1)},\dots,\pmb{J}_{\alpha_{ji,b}}^{(N_M)}\}$, and
$\pmb{J}_{\alpha_{ji,b}}^{(k)}=[\text{Re}\{\pmb{F}_{L_i,b}^*\pmb{F}_{L_i,b}\}\,\,-\text{Im}\{\pmb{F}_{L_i,b}^*\pmb{F}_{L_i,b}\};$ $\text{Im}\{\pmb{F}_{L_i,b}^*$ $\pmb{F}_{L_i,b}\}\,\,\text{Re}\{\pmb{F}_{L_i,b}^*\pmb{F}_{L_i,b}\}]$ for $k \in \mathcal{N_M}$. After some algebra, the FIM $\pmb{J}_{\pmb{\theta}_i}$ can be again written as in (\ref{eq:reFIM}), where 
$\pmb{A}=\sum_{j \in \mathcal{N}_B}\sum_{b \in \mathcal{N}_c} \pmb{G}_{ji}\pmb{\Psi}_{ji,b}\pmb{G}_{ji}^T=\sum_{j\in \mathcal{N}_B}\sum_{b\in \mathcal{N}_c}$ $\frac{2}{c^2N_0} \pmb{J}_{\tau_{ji,b}}\pmb{q}_{ji}\pmb{q}_{ji}^T$, $\pmb{B}=[\pmb{G}_{1i}\pmb{\Psi}_{1i,1}\pmb{D}^T$ $\cdots\pmb{G}_{N_Bi}\pmb{\Psi}_{N_Bi,N_C}\pmb{D}^T]
=\frac{2}{cN_0}[\pmb{q}_{1i}\pmb{J}_{1i,1}^T\cdots\pmb{q}_{N_Bi}\pmb{J}_{N_Bi,N_C}^T]$ and $\pmb{C}=\text{diag}\{\pmb{D}$ $\pmb{\Psi}_{1i,1}\pmb{D}^T,\dots, \pmb{D}\pmb{\Psi}_{N_Bi,N_C}\pmb{D}^T\}
=\frac{2}{N_0}\text{diag}\{\pmb{J}_{\alpha_{1i,1}},\dots,\pmb{J}_{\alpha_{N_Bi,N_C}}\}$, where $\pmb{D}=[\pmb{0}_{2N_ML_i}\,\,\pmb{I}_{2N_ML_i}]$. By applying the Schur complement (see, e.g., \cite{Zhang05Book}), we can finally obtain the EFIM as (\ref{FIMTOAselec}).

\bibliographystyle{IEEEtran}
\bibliography{JSAref}

\begin{thebibliography}{10}
\providecommand{\url}[1]{#1}
\csname url@samestyle\endcsname
\providecommand{\newblock}{\relax}
\providecommand{\bibinfo}[2]{#2}
\providecommand{\BIBentrySTDinterwordspacing}{\spaceskip=0pt\relax}
\providecommand{\BIBentryALTinterwordstretchfactor}{4}
\providecommand{\BIBentryALTinterwordspacing}{\spaceskip=\fontdimen2\font plus
\BIBentryALTinterwordstretchfactor\fontdimen3\font minus
  \fontdimen4\font\relax}
\providecommand{\BIBforeignlanguage}[2]{{%
\expandafter\ifx\csname l@#1\endcsname\relax
\typeout{** WARNING: IEEEtran.bst: No hyphenation pattern has been}%
\typeout{** loaded for the language `#1'. Using the pattern for}%
\typeout{** the default language instead.}%
\else
\language=\csname l@#1\endcsname
\fi
#2}}
\providecommand{\BIBdecl}{\relax}
\BIBdecl

\bibitem{JSAGlobalSIP13}
S.~Jeong, O.~Simeone, A.~Haimovich, and J.~Kang, ``Beamforming design for joint
  localization and data transmission in distributed antenna system,'' in
  \emph{Proc. IEEE GlobalSIP}, Austin, TX, 3-5 Dec. 2013, pp. 879--882.

\bibitem{Makela_02Mag}
K.~Pahlavan, X.~Li, and J.-P. Makela, ``Indoor geolocation science and
  technology,'' \emph{IEEE Commun. Magazine}, vol.~40, no.~2, pp. 112--118,
  Feb. 2002.

\bibitem{Kumar03spmag}
C.-Y. Chong and S.~P. Kumar, ``Sensor networks: evolution, opportunities, and
  challenges,'' \emph{Proceedings of the IEEE}, vol.~91, no.~8, pp. 1247--1256,
  Aug. 2003.

\bibitem{Gezici_05spmag}
S.~Gezici, Z.~Tian, G.~B. Giannakis, H.~Kobayashi, A.~F. Molisch, H.~V. Poor,
  and Z.~Sahinoglu, ``Localization via ultra-wideband radios: a look at
  positioning aspects for future sensor networks,'' \emph{IEEE Signal Proc.
  Magazine}, vol.~22, no.~4, pp. 70--84, July 2005.

\bibitem{Guvenc09CST}
I.~Guvenc and C.-C. Chong, ``A survey on {TOA} based wireless localization and
  {NLOS} mitigation techniques,'' \emph{IEEE Commun. Surveys \& Tutorials},
  vol.~11, no.~3, pp. 107--124, 3rd Quarter 2009.

\bibitem{Shen12Globecom}
Y.~Shen, W.~Dai, and M.~Z. Win, ``Optimal power allocation for active and
  passive localization,'' in \emph{Proc. IEEE GLOBECOM 2012}, Anaheim, CA, 3-7
  Dec. 2012, pp. 3713--3718.

\bibitem{Shen13ACM}
Y.~Shen, W.~Dai, and M.~Z. Win, ``Power optimization for network localization,'' \emph{IEEE/ACM Trans.
  on Networking}, vol.~PP, no.~99, pp. 1--14, Sep. 2013.

\bibitem{Li13Arxiv}
W.~W.-L. Li, Y.~Shen, Y.~Jun, and M.~Z. Win, ``Robust power allocation for
  energy-efficient location aware networks,'' \emph{arXiv preprint
  arXiv:1305.6091}, 2013.

\bibitem{Wang13SigJ}
T.~Wang, Y.~Shen, S.~Mazuelas, H.~Shin, and M.~Z. Win, ``On {OFDM} ranging
  accuracy in multipath channels,'' \emph{IEEE Systems Journal}, vol.~PP,
  no.~99, pp. 1--11, Oct. 2013.

\bibitem{Antique97VIRJIN}
M.~Aatique, ``Evaluation of {TDOA} techniques for position location in {CDMA}
  systems,'' Ph.D. dissertation, Virginia Polytechnic Institute and State
  University, 1997.

\bibitem{Montal13ICASSP}
R.~Montalban, J.~A. L{\'o}pez-Salcedo, G.~Seco-Granados, and A.~L.
  Swindlehurst, ``Power allocation method based on the channel statistics for
  combined positioning and communications {OFDM} systems,'' in \emph{Proc. IEEE
  ICASSP 2013}, Vancouver, Canada, May 2013, pp. 4384--4388.

\bibitem{Montal12ASIL}
R.~Montalban, G.~Seco-Granados, and A.~L. Swindlehurst, ``Suboptimal method for
  pilot and data power allocation in combined positioning and communications
  {OFDM} systems,'' in \emph{Proc. IEEE ASILOMAR 2012}, Pacific Grove, CA, 4-7
  Nov. 2012, pp. 1041--1045.

\bibitem{Larsen11ICASSP}
M.~D. Larsen, G.~Seco-Granados, and A.~L. Swindlehurst, ``Pilot optimization
  for time-delay and channel estimation in {OFDM} systems,'' in \emph{Proc.
  IEEE ICASSP 2011}, Prague Congress Centre Prague, Czech Republic, 22-27 May
  2011, pp. 3564--3567.

\bibitem{Kay93Book}
S.~M. Kay, \emph{Fundamentals of Signal Processing-Estimation Theory}.\hskip
  1em plus 0.5em minus 0.4em\relax Englandwood Cliffs, NJ: Prentice Hall, 1993.

\bibitem{Luo10SPM}
Z.-Q. Luo, W.-K. Ma, A.-C. So, Y.~Ye, and S.~Zhang, ``Semidefinite relaxation
  of quadratic optimization problems,'' \emph{IEEE Signal Proc. Magazine},
  vol.~27, no.~3, pp. 20--34, May 2010.

\bibitem{Beng99Allerton}
M.~Bengtsson and B.~Ottersten, ``Optimal downlink beamforming using
  semidefinite optimization,'' in \emph{Proc. Allerton Conf. on Commun. Control
  and Computing}, vol.~37, Monticello, IL, Sep. 1999, pp. 987--996.

\bibitem{Beck10Book}
{A. Beck} and {M. Teboulle}, \emph{Gradient-based algorithms with applications
  to signal recovery}.\hskip 1em plus 0.5em minus 0.4em\relax in Convex
  Optimization in Signal Processing and Communications, Edited by Daniel P.
  Palomar and Yonina C. Eldar, Cambridge University Press, 2009.

\bibitem{Ben09Book}
A.~Ben-Tal, L.~El~Ghaoui, and A.~Nemirovski, \emph{Robust optimization}.\hskip
  1em plus 0.5em minus 0.4em\relax Princeton University Press, 2009.

\bibitem{Molit05Book}
A.~F. Molisch, \emph{Wireless communications}.\hskip 1em plus 0.5em minus
  0.4em\relax Wiley, 2010.

\bibitem{Sam03TSP}
D.~Samardzija and N.~Mandayam, ``Pilot-assisted estimation of {MIMO} fading
  channel response and achievable data rates,'' \emph{IEEE Trans. Signal
  Proc.}, vol.~51, no.~11, pp. 2882--2890, Nov. 2003.

\bibitem{Shen10TIT}
Y.~Shen and M.~Z. Win, ``Fundamental limits of wideband localization—part
  {I}: a general framework,'' \emph{IEEE Trans. Inform. Theory}, vol.~56,
  no.~10, pp. 4956--4980, Oct. 2010.

\bibitem{Heath13Mag}
R.~Heath, S.~Peters, Y.~Wang, and J.~Zhang, ``A current perspective on
  distributed antenna systems for the downlink of cellular systems,''
  \emph{IEEE Commun. Magazine}, vol.~51, no.~4, pp. 161--167, Apr. 2013.

\bibitem{STDLTE3GPP}
{3GPP} {TS} 36.101, "{E-UTRA} {UE} radio transmission and reception," (Release
  10), V10.3.0.

\bibitem{STDE911}
D.~N. Hateld, ``{A} report on technical and operational issues impacting the
  provision of wireless enhanced 911 services,'' {T}echnical report, Federal
  Communications Commission, 2002.

\bibitem{Zhang05Book}
F.~Zhang, \emph{The Schur complement and its applications}.\hskip 1em plus
  0.5em minus 0.4em\relax Springer, 2005, vol.~4.

\bibitem{Hunter04Amer}
D.~R. Hunter and K.~Lange, ``A tutorial on {MM} algorithms,'' \emph{The
  American Statistician}, vol.~58, no.~1, pp. 30--37, Feb. 2004.

\bibitem{Boyd04Book}
S.~P. Boyd and L.~Vandenberghe, \emph{Convex optimization}.\hskip 1em plus
  0.5em minus 0.4em\relax Cambridge university press, 2004.

\end{thebibliography}
\end{document}